\documentclass[pre,onecolumn,preprintnumbers,floatfix,amssymb,amsmath]{revtex4}
\usepackage{graphicx}
\usepackage{subfigure}

\begin{document}
\title{Self--assembly of a drop pattern from a two-dimensional grid of nanometric metallic filaments}
\author{Ingrith Cuellar}

\author{Pablo D. Ravazzoli}

\author{Javier A. Diez}

\author{Alejandro G. Gonz\'alez}\email{aggonzal@exa.unicen.edu.ar}
\affiliation{Instituto de F\'{\i}sica Arroyo Seco, Universidad Nacional del Centro de la Provincia de Buenos Aires and CIFICEN-CONICET-CICPBA, Pinto 399, 7000, Tandil, Argentina}

\author{Nicholas A. Roberts}
\affiliation{Mechanical and Aerospace Engineering, Utah State University, Logan, Utah 84322, USA}

\author{Jason D. Fowlkes}
\affiliation{Center for Nanophase Materials Sciences, Oak Ridge National Laboratory, Oak Ridge, Tennessee 37381, USA}
\affiliation{Department of Materials Science \& Engineering, University of Tennessee, Knoxville, Tennessee 37996, USA}

\author{Philip D. Rack}
\affiliation{Center for Nanophase Materials Sciences, Oak Ridge National Laboratory, Oak Ridge, Tennessee 37381, USA}
\affiliation{Department of Materials Science \& Engineering, University of Tennessee, Knoxville, Tennessee 37996, USA}

\author{Lou Kondic}
\affiliation{Department of Mathematical Sciences, New Jersey Institute of Technology, Newark, New Jersey 07102, USA}

\begin{abstract}
We report experiments, modeling and numerical simulations of the self--assembly of particle patterns obtained from a  nanometric metallic square grid. Initially, nickel filaments of rectangular cross section are patterned on a SiO$_2$ flat surface, and then they are melted by laser irradiation with $\sim 20$~ns pulses. During this time, the liquefied metal dewets the substrate, leading to a linear array of drops along each side of the squares. The experimental data provides a series of SEM images of the resultant morphology as a function of the number of laser pulses or cumulative liquid lifetime. These data are analyzed in terms of fluid mechanical  models that account for mass conservation and consider flow evolution with the aim to predict the final number of drops resulting from each side of the square. The aspect ratio, $\delta$, between the square sides' lengths and their widths is an essential parameter of the problem. Our models allow us to predict the $\delta$--intervals within which a certain final number of drops are expected. The comparison with experimental data shows a good agreement with the model that explicitly considers the Stokes flow developed in the filaments neck region that lead to breakup points. Also, numerical simulations, that solve the Navier--Stokes equations along with slip boundary condition at the contact lines, are implemented to describe the dynamics of the problem.

\end{abstract}

\maketitle

\section{Introduction}

Controlling the placement and size of metallic nanostructures is crucial in many applications~\cite{ruffino_pssa15}. For instance, the surface plasmon resonance among metallic drops depends on a coordination between their size and spacing~\cite{halas_cr11,le_acsn08}. Thus, the inclusion of this type of nanoparticles into photovoltaic devices has led to increased efficiency~\cite{atwater_natmat10,wu_acsn11}. Moreover, in the field of biodiagnostics and sensing, functionalized Au nanometric drops bind to specific DNA markers thus permitting binding detection~\cite{rosi_cr05}. In general, the potential applications of organized metallic nanostructures are wide--ranging and include Raman spectroscopy~\cite{anker_nm08,vo-dinh_trac98}, catalysis~\cite{christopher_ar11}, photonics~\cite{ozbay_sc06} and spintronics~\cite{wolf_sc01}. A methodology to generate and organize structures at the nanoscale is to take advantage of the natural tendency of materials to the self--assemble~\cite{honisch_pssa15,koplik_pof06}. By combining the fact that liquid metals have low viscosity and high surface energy with the current highly developed nanoscale lithography techniques, we have a platform to study the governing liquid--state dewetting dynamics~\cite{lian_nanol06,dgk_pof09} such as liquid instabilities~\cite{wu_lang11} with the goal of directing the assembly of precise, coordinated nanostructures in one~\cite{fowlkes_nl14} and two~\cite{roberts_acsami13} dimensions.

In this work, we focus on the formation of a two--dimensional drop pattern starting from the pulsed laser--induced dewetting (PliD)~\cite{roberts_acsami13} of a square grid of Ni strips on a SiO$_2$ coated silicon wafer. To investigate the behavior of these melted square grids, we employ well established nanofabrication techniques and PliD. With this methodology it is experimentally possible to precisely control the initial far--from-equilibrium geometry and the liquid lifetime via nanosecond laser melting. 

Fluid dynamics is used to rationalize the experimental data, because the evolution (and instability) of the metal shape occurs in liquid state, and we therefore focus on its analysis from the fluid dynamical point of view.  For simplicity, we consider the liquid metal as a Newtonian fluid, and ignore the effects that evolving metal temperature has on the material properties. The models that are discussed are based on 
earlier ones developed in the context of experiments involving the evolution of grids made of silicon oil filaments. While the scale of the experiments considered in the current work is considerably smaller, we will see that the main modeling approaches developed for the films of millimetric thickness are useful to describe the results on nanoscale as well. 

The metal geometry analyzed here, while related to the ones studied in previous works involving nanoscale metal films~\cite{roberts_acsami13,fowlkes_nl14,hartnett_lang15} provides new and interesting challenges and open questions. The process of breakup of an original grid into filaments is of interest on its own. For example, one could ask whether a drop will form at the intersection points, or whether a dry spot will be present there? Does the answer depend on the deposited film filament? Once the independent filaments form, can their evolution be described based on the stability analysis of an infinite cylinder?  And finally, to which degree could the results be explained based on fluid mechanical models and simulations of Navier-Stokes equations?  

We will discuss in the following section the complex procedure involving heating, melting, and consequent solidification of liquid metal filaments. While significant amount of previous work~\cite{wu_lang11,gonzalez_lang13} suggests that focusing on simple isothermal Newtonian formulations leads to a reasonable agreement with experiments, it is not a priori clear that such an approach is appropriate for a rather complex geometry of metal grids/meshes on the nanoscale.  For example, some works suggest that thermal gradients leading to Marangoni effect may be relevant~\cite{oron_pof98,oron_pof00,Trice_07,trice_prl08}, although recent work focusing on liquid filaments suggests that in this geometry they could be safely ignored~\cite{seric_pof18}.   

This paper is organized as follows.  In the experimental section we give a brief outline of the setup. In following section, results are presented and we use three models to describe the instability observed in experiments. We start from the conceptually simplest linear stability analysis, and proceed to consider progressively more complicated models based on mass conservation (MCM) and on a fluid dynamic model (FDM) for the evolution of the breakups.  Next, we report Navier--Stokes numerical simulations using an appropriate geometry based on the experiments and we discuss some special effects observed at the grid corners. Finally, we summarize the results and consider future perspectives.

\section{Experimental Section}

Electron beam lithography followed by direct current magnetron sputtering is used to define square grids of Ni strips on Si wafers (coated with a  $100$~nm thickness SiO$_2$ layer), which were later melted by nanosecond laser pulses.  In this section, we describe the details of the experimental procedure.

{\it Electron beam lithography:} Focused electron beam exposure at $100 $ keV and $2$  nA was conducted using a JEOL 9300 electron beam lithography system on poly(methylmethacrylate) (PMMA, positive tone electron sensitive resist 495-A4 provided by Shipley) in order to define the strips that will form the square grid. The PMMA resist was previously spin coated on a $100$~mm diameter substrate rotated at $4000$  rpm during $45$ s. The spin coating process was followed by a $2$ min, $180^\circ$C hot plate bake. An electron beam dose of $1000 \, \mu {\rm C} \, \rm{cm}^{-2}$ was required in order to completely expose the electron resist yielding well defined the thin film strips of the grid. A 495-A4 resist development was carried out in a $1$:$3$ methyl isobutyl ketone (MIBK)--isopropyl alcohol (IPA) solution during $100$ s followed by an IPA rinse in order to expose the strips in the resist down to the underlying SiO$_2$ layer. Any residual electron resist was removed by exposing the system to an oxygen plasma generated in a reactive ion etcher  for $8$  s ($100$ W capacitively coupled plasma, $10~\rm{cm}^3\ \rm{min}^{-1}$ O$_2$ flow rate and a pressure setting of $150$ mTorr). 

{\it DC magnetron sputtering deposition:} An AJA International $200$ DC magnetron sputtering system was used to deposit the Ni thin film strips. The process was carried on with a constant power deposition mode at $30$ W at a chamber pressure of $3$ \ mTorr Ar, which was maintained using a gas flow rate of 
$25 \, \rm{cm}^3 \, \rm {min}^{-1}$. The sputter rate of Ni was $5.8$ nm min$^{-1}$ for a target--to--substrate distance of $~5$ cm. A wet, metal lift--off procedure consisting of the immersion of the substrate chip in acetone for $1$~min was used to dissolve unexposed resist. Thus, the lift--off of the unwanted metal layer surrounding the Ni thin film strip features was achieved. Subsequently, the  substrate chip was rinsed in acetone, afterwards in isopropyl alcohol, and finally blown dry using $N_2$ gas to remove any remaining debris from the substrate. No specific treatments were done to remove the native Ni oxide prior to laser irradiation, as no obvious influences of the native oxide have been observed in the assembly dynamics.

{\it Nanosecond, ultraviolet pulsed laser irradiation:} A Lambda Physik LPX--$305i$, KrF excimer laser ($248$ nm wavelength) was used to irradiate and melt the Ni square grid. During irradiation, the substrate surface was normal to the incident laser pulse. As a result, the top surface of the strips as well as the surrounding substrate surface were irradiated. The incident beam size was on the order of $\sim 1$~cm$^2$, significantly larger than the grid area ($\sim \mu \rm{m}^2$), and thus irradiated the grids in a uniform way. The pulse width of the laser beam was $\sim 18$ ns (FWHM). A beam fluence of $(200 \pm 10)$ mJ  cm$^{-2}$ was used to melt the grid, and focusing of the output beam was required in order to achieve this fluence. All samples reported in this work were irradiated with (at most) $30$ laser pulses. 

The Ni square grids were patterned with rectangular cross section strips of width $w_g=(162\pm 6)$~nm, and length $L_g$ (internal side of the grid squares) in the range $(600,1800)$~nm. We consider three different thicknesses, namely  $h_g=5$, $10$, and $20$~nm ($\pm 1$~nm).  

\section{Results and Discussion}

\begin{figure}[hbt]
\subfigure[\,Initial state]
{\includegraphics[width=0.325\linewidth]{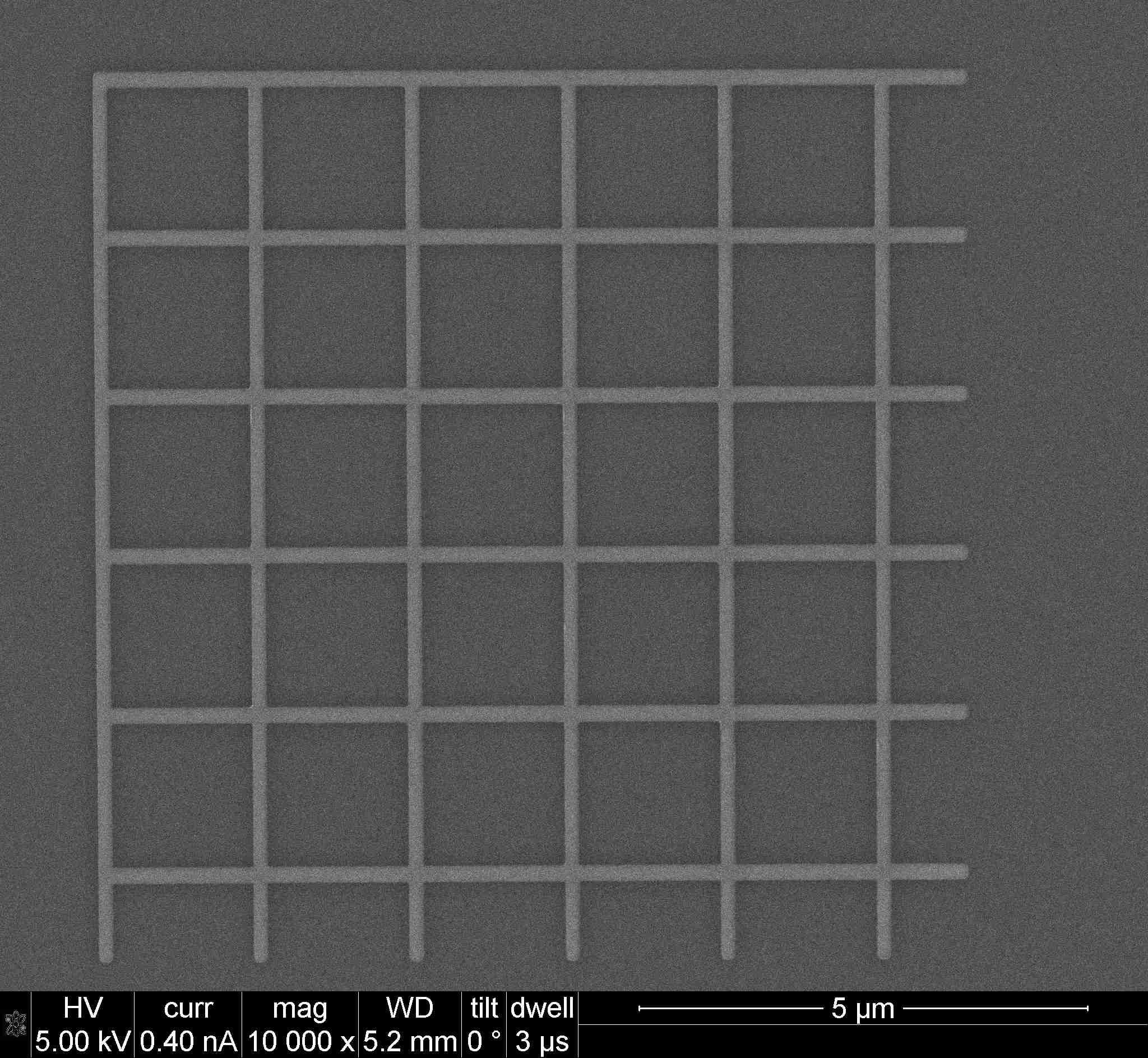}}
\subfigure[\,$1^{st}$ pulse]
{\includegraphics[width=0.325\linewidth]{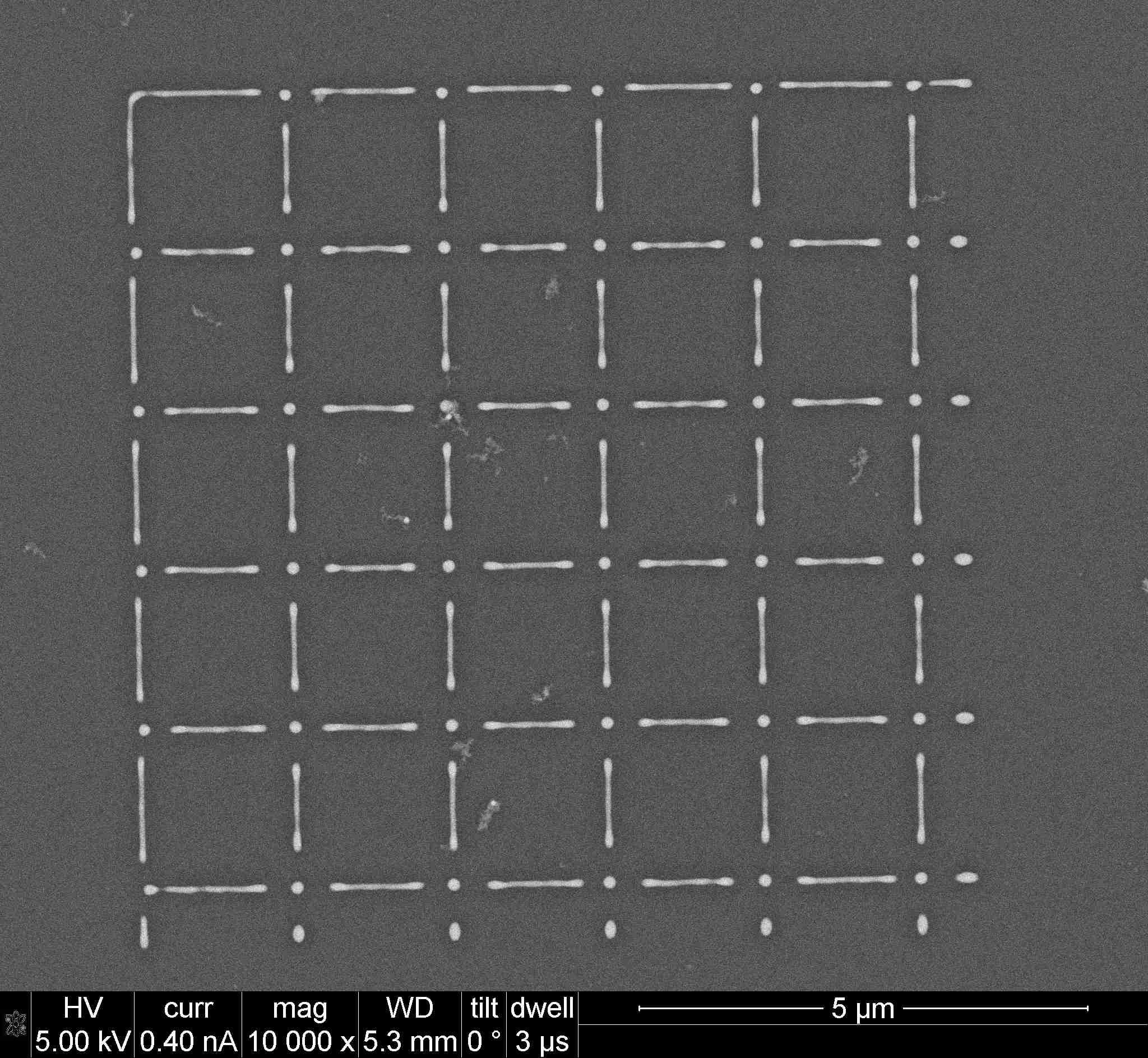}}
\subfigure[\,$5^{th}$ pulse]
{\includegraphics[width=0.325\linewidth]{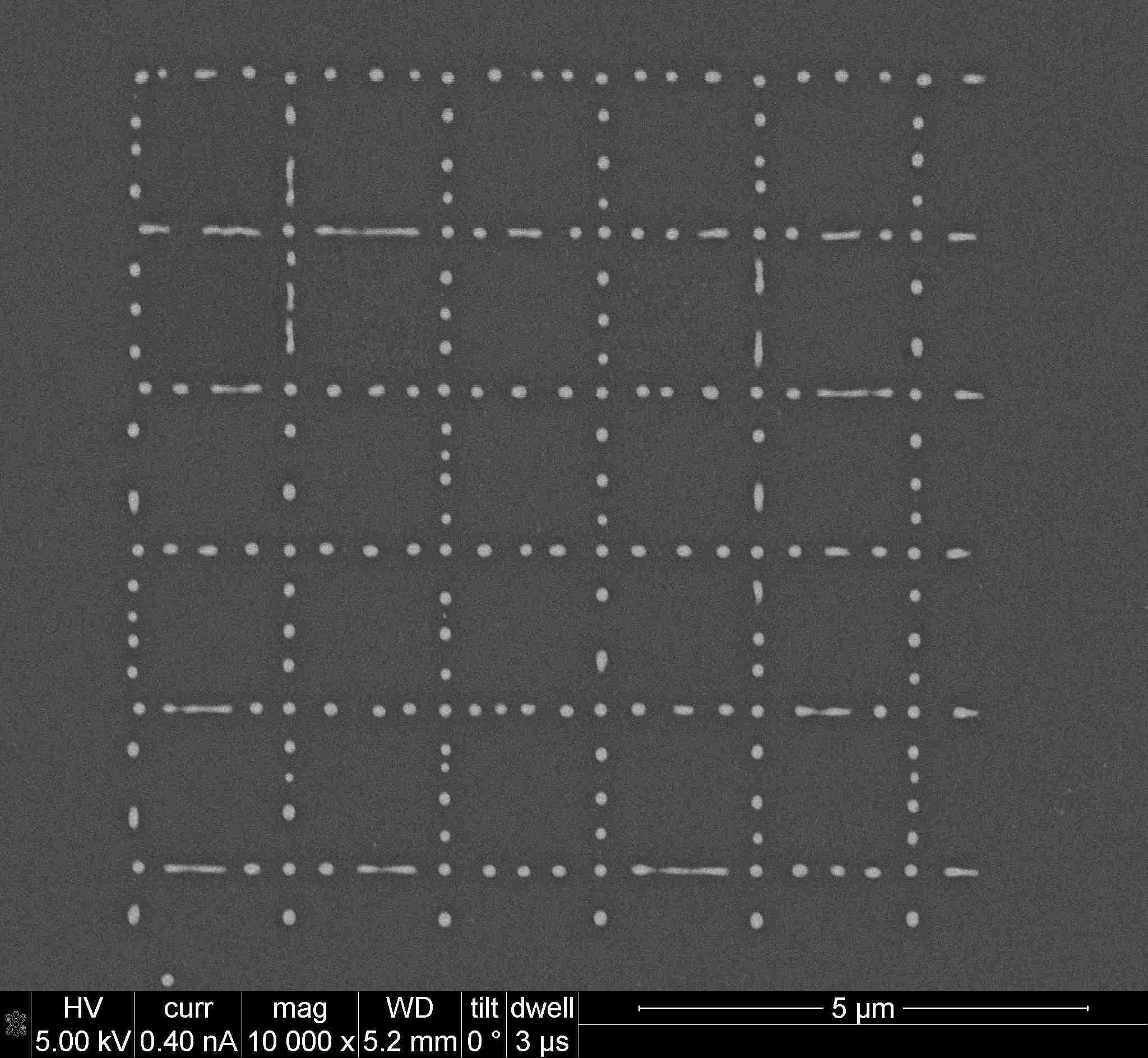}}
\caption{(a) Initial square grid of Ni strips with rectangular cross section. The as-deposited metal thickness is $h_g=(10 \pm 1)$~nm. The inner square sides are $L_g=1587$~nm long. (b) After the first laser pulse, the strips decrease their width by dewetting and the ends detach from the vertices, leaving drops there. (c) Drops pattern resulting from the breakup of the filaments after $5$ pulses.}
\label{fig:grid_001}
\end{figure}

A typical example of the initial state is shown in Figure~\ref{fig:grid_001}a, which corresponds to $h_g=10$~nm and $L_g=1587$~nm. At the end of the first pulse, a single drop appears at the vertices, while shorter and narrower filaments with small bulges at the ends are formed along the sides of the squares (see Fig.~\ref{fig:grid_001}b). This structure is a consequence of a liquid--like behavior of the strips due to melting. For subsequent pulses, axial retractions from both ends (shortening) of the remaining filaments are observed in an iterative fashion.  The resulting further bulges with the corresponding new bridges lead to the formation of a certain number of drops along the sides of the squares. Figure~\ref{fig:grid_001}c shows the pattern obtained after $5$ pulses, when the evolution has almost finished. 

Once the initial strip has been melted, its rectangular cross section evolves into a cylindrical cap shape by parallel contact line retractions (dewetting) thus leading to a narrower filament of width
\begin{equation}
 w= 2 \sqrt{\frac{h_g w_g}{\theta-\sin \theta \cos \theta}} \sin \theta,
 \label{eq:width}
\end{equation}
where $\theta$ is the static contact angle. For the present Ni/SiO$_2$ system, we have $\theta=69^\circ \pm 8^\circ$~\cite{fowlkes_nl14}.  Thus, we have 
\begin{equation}
w=(2 \pm 0.17) \sqrt{h_g w_g}.  
\label{eq:w69}
\end{equation}
A comparison between the measured widths and the calculated values given by Eq.~(\ref{eq:width}) shows a very good agreement. For instance, for $h_g=10$~nm, we have measured an average width of $(78.2 \pm 3.6)$~nm, while the calculated value is $w=(80.7 \pm 7)$~nm.

\begin{figure}[hbt]
\subfigure[\,$L_g=1387$~nm ($\delta=17.21$)]
{\includegraphics[width=0.325\linewidth]{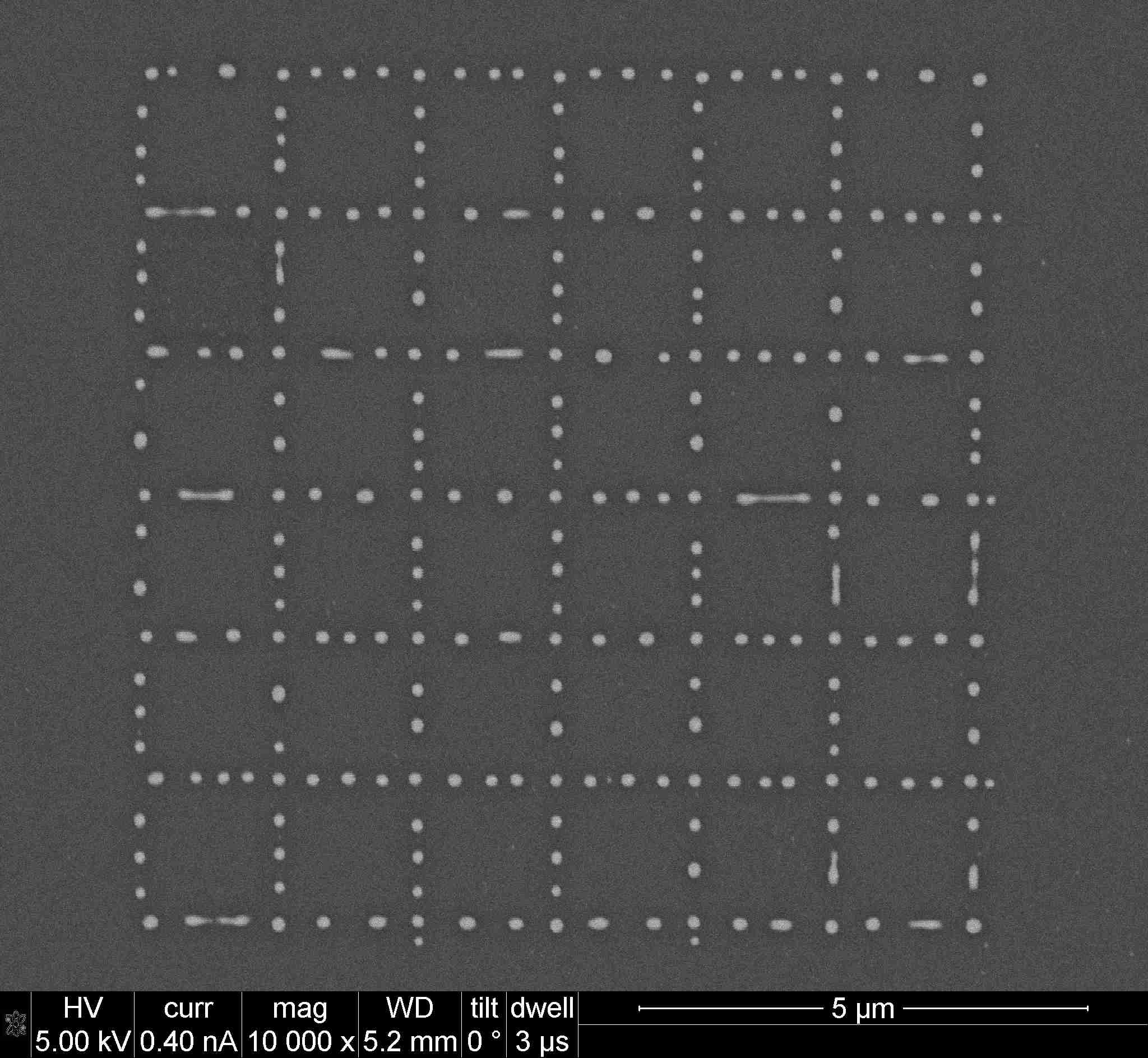}}
\subfigure[\,$L_g=987$~nm ($\delta=12.25$)]
{\includegraphics[width=0.325\linewidth]{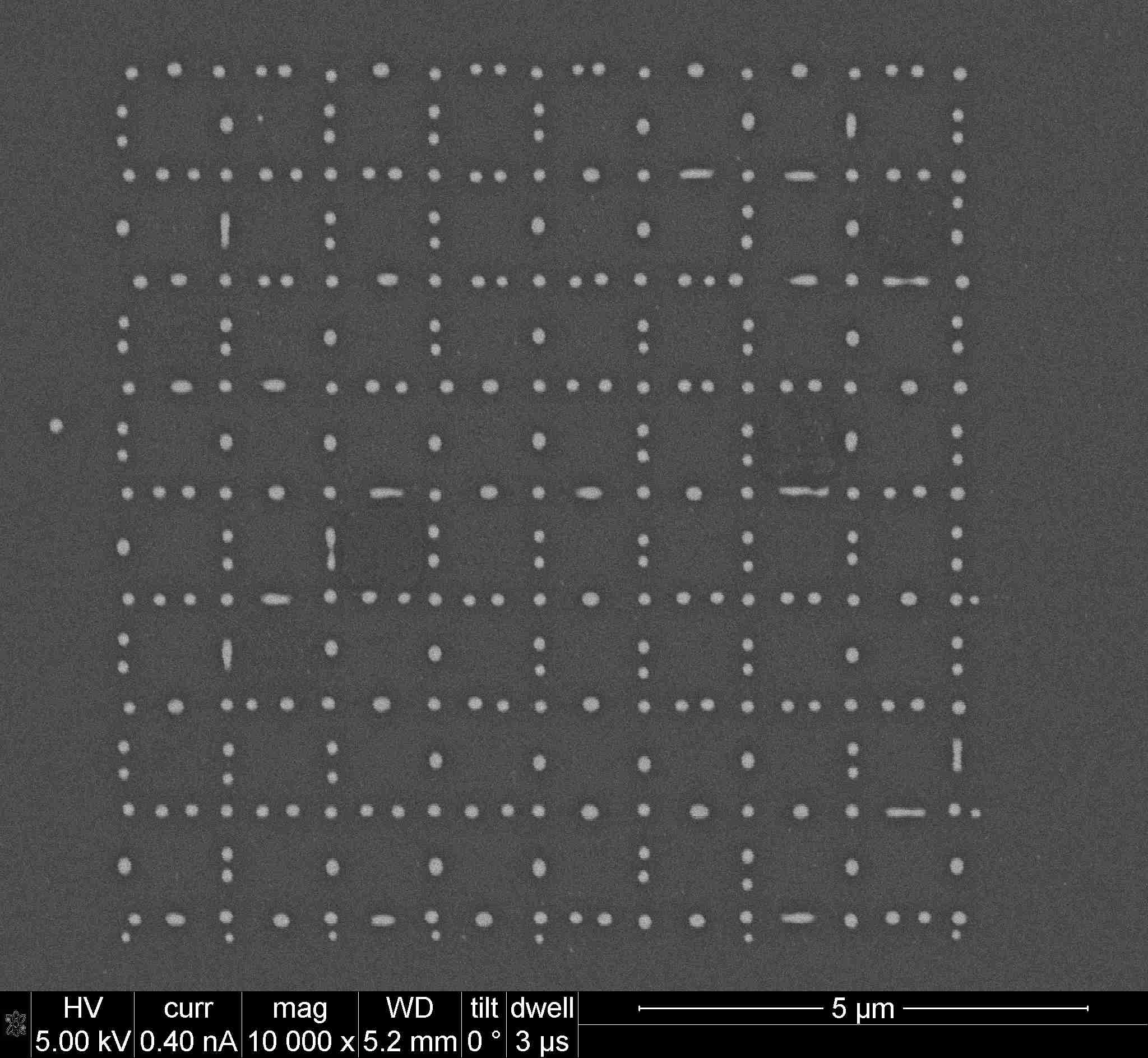}}
\subfigure[\,$L_g=606$~nm ($\delta=7.52$)]
{\includegraphics[width=0.325\linewidth]{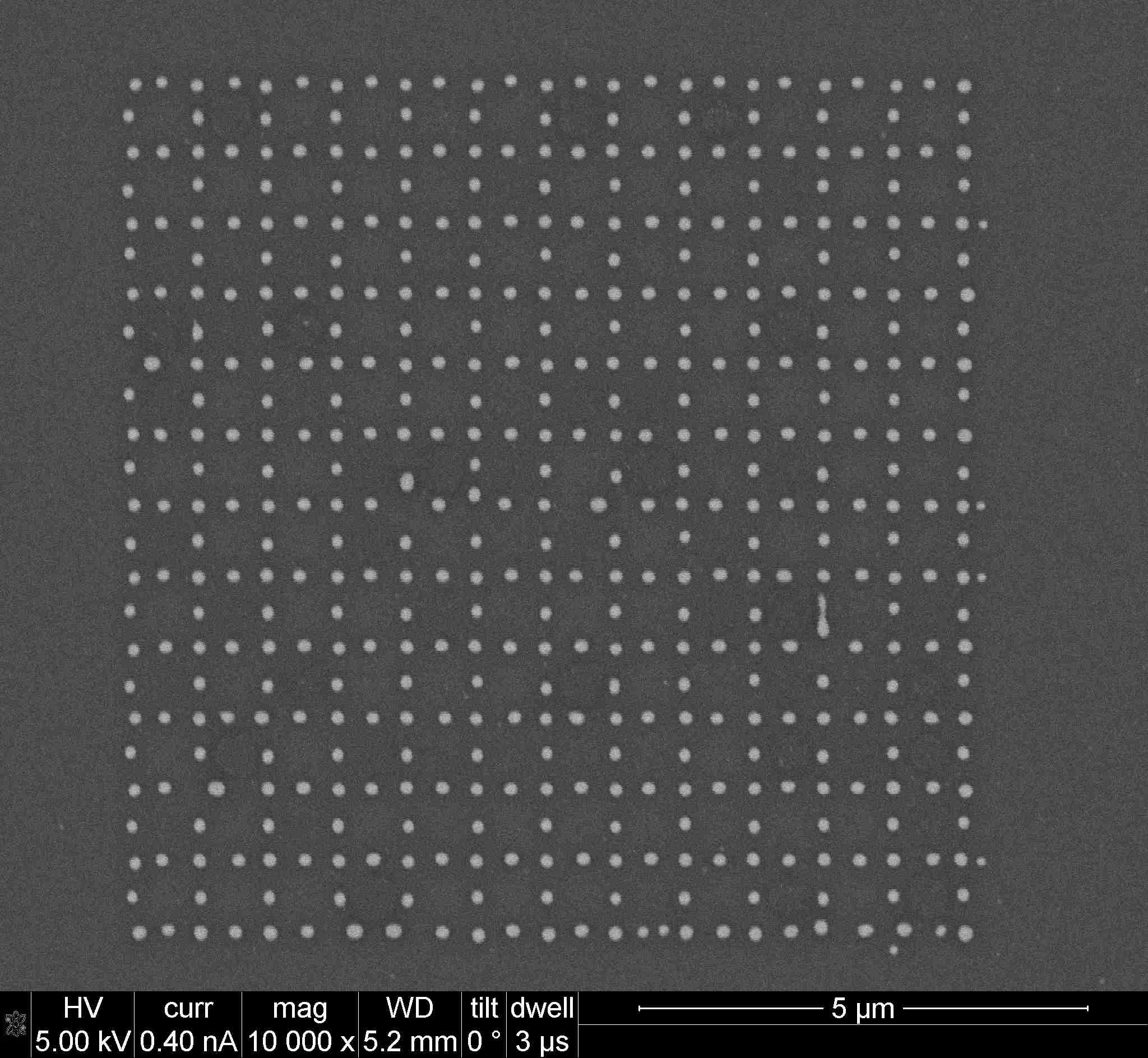}}
\caption{(a) Drop patterns for $h_g=10$~nm and $w_g=162$~nm after $5$ pulses for different initial lengths, $L_g$.}
\label{fig:grid_Lg}
\end{figure}

Figure~\ref{fig:grid_Lg} shows the patterns observed after $5$ pulses for three values of $L_g$, with $h_g=10$~nm. Clearly, the number of drops, $n$, along each side, decreases with $L_g$ (note that $n$ does not include the drops at the vertices). Even if there is a dominant value of $n$ in each case, some dispersion of $n$ is observed. 
We expect that the main reason for this dispersion is the experimental noise leading to edge roughness of the strips. Figures~\ref{fig:grid_Lg}(a) and (b) show that after $5$ pulses there are still few filaments which have not yet finished their breakup. Additional pulses will lead to full particle formation, but since there are only few we ignore them, and consider in our analysis only the filaments that have finished evolving.  

\begin{figure}[htb]
\subfigure[$\,h_g=5$~nm]
{\includegraphics[width=0.32\linewidth]{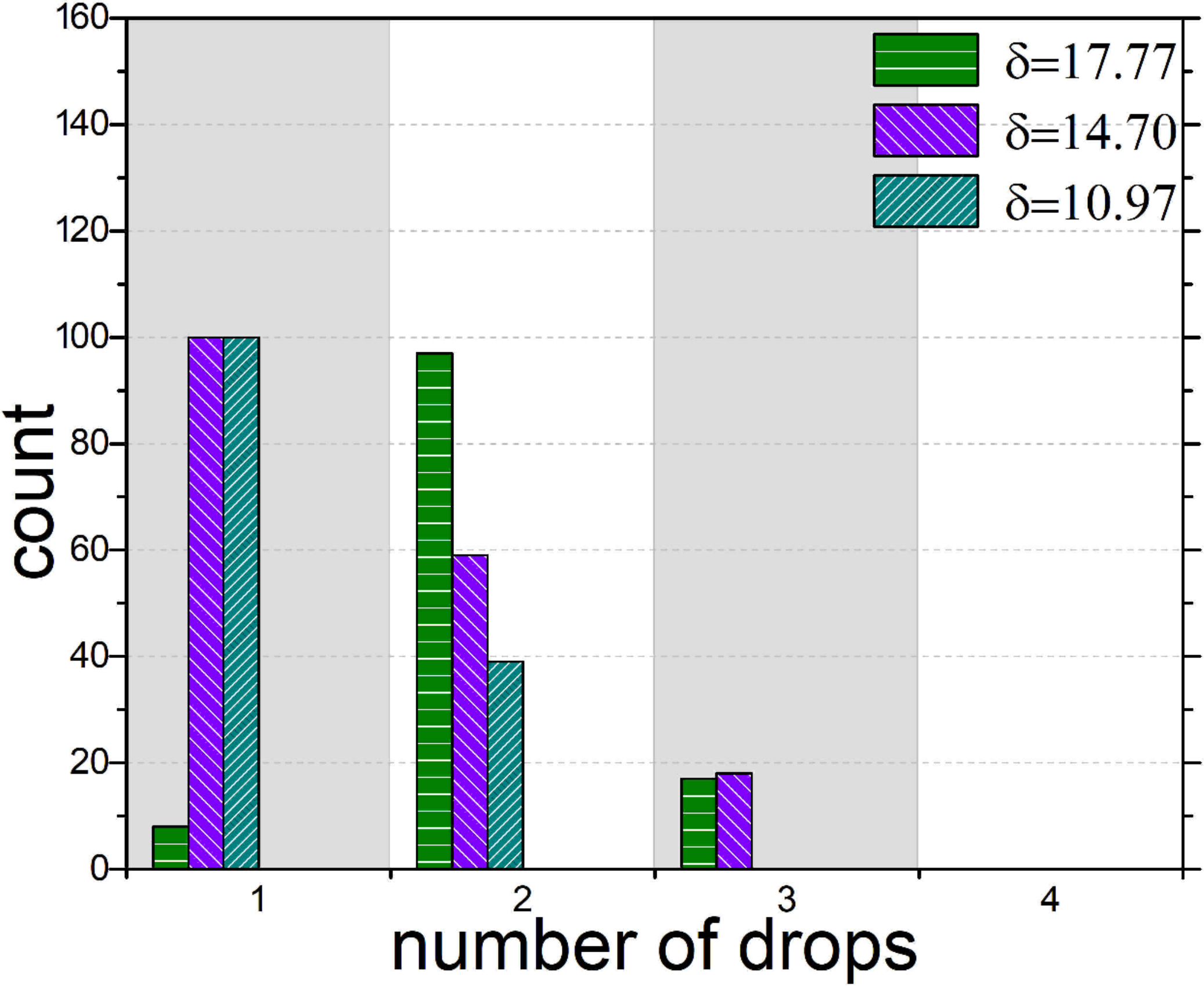}}
\subfigure[$\,h_g=10$~nm]
{\includegraphics[width=0.32\linewidth]{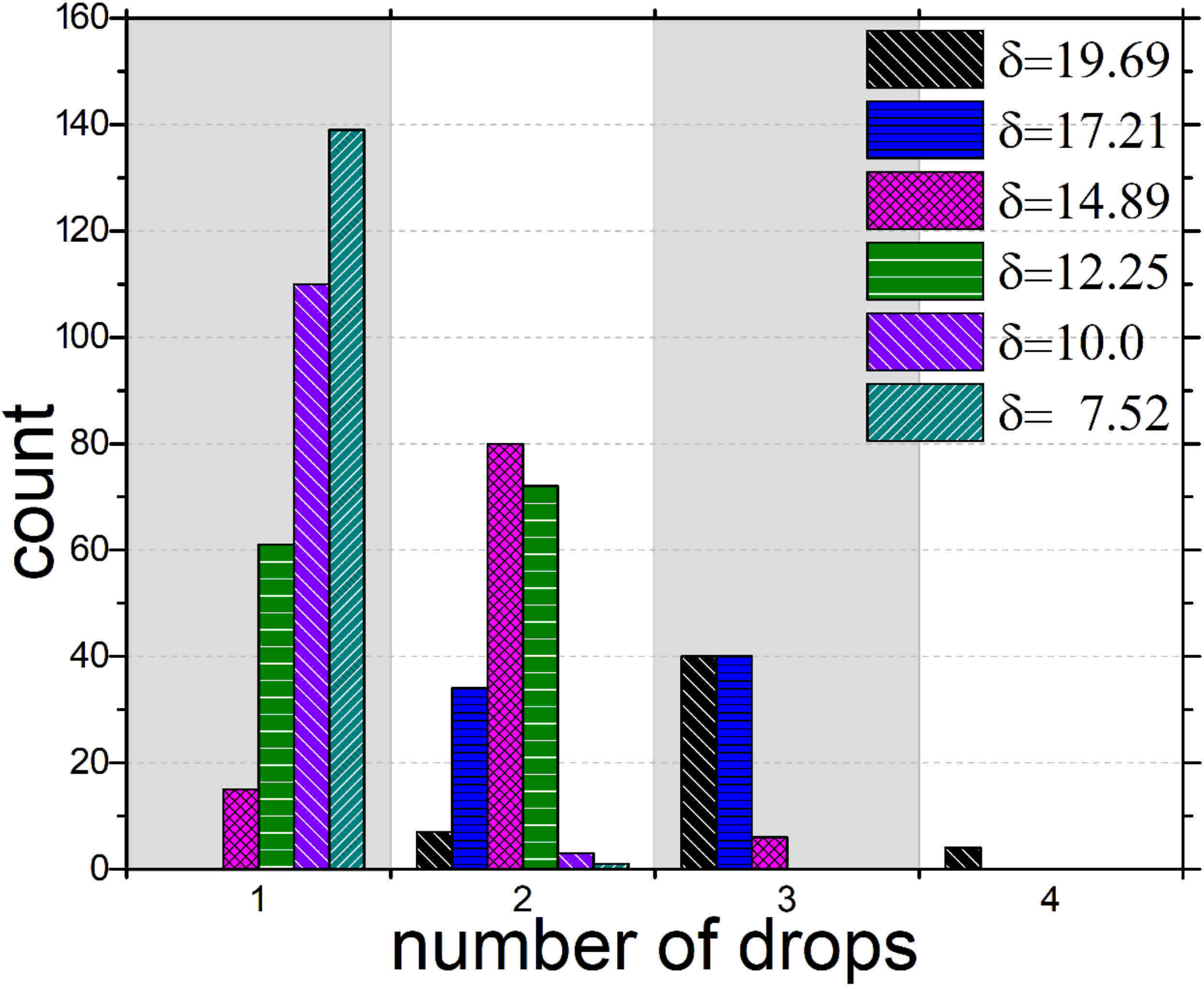}}
\subfigure[$\,h_g=20$~nm]
{\includegraphics[width=0.32\linewidth]{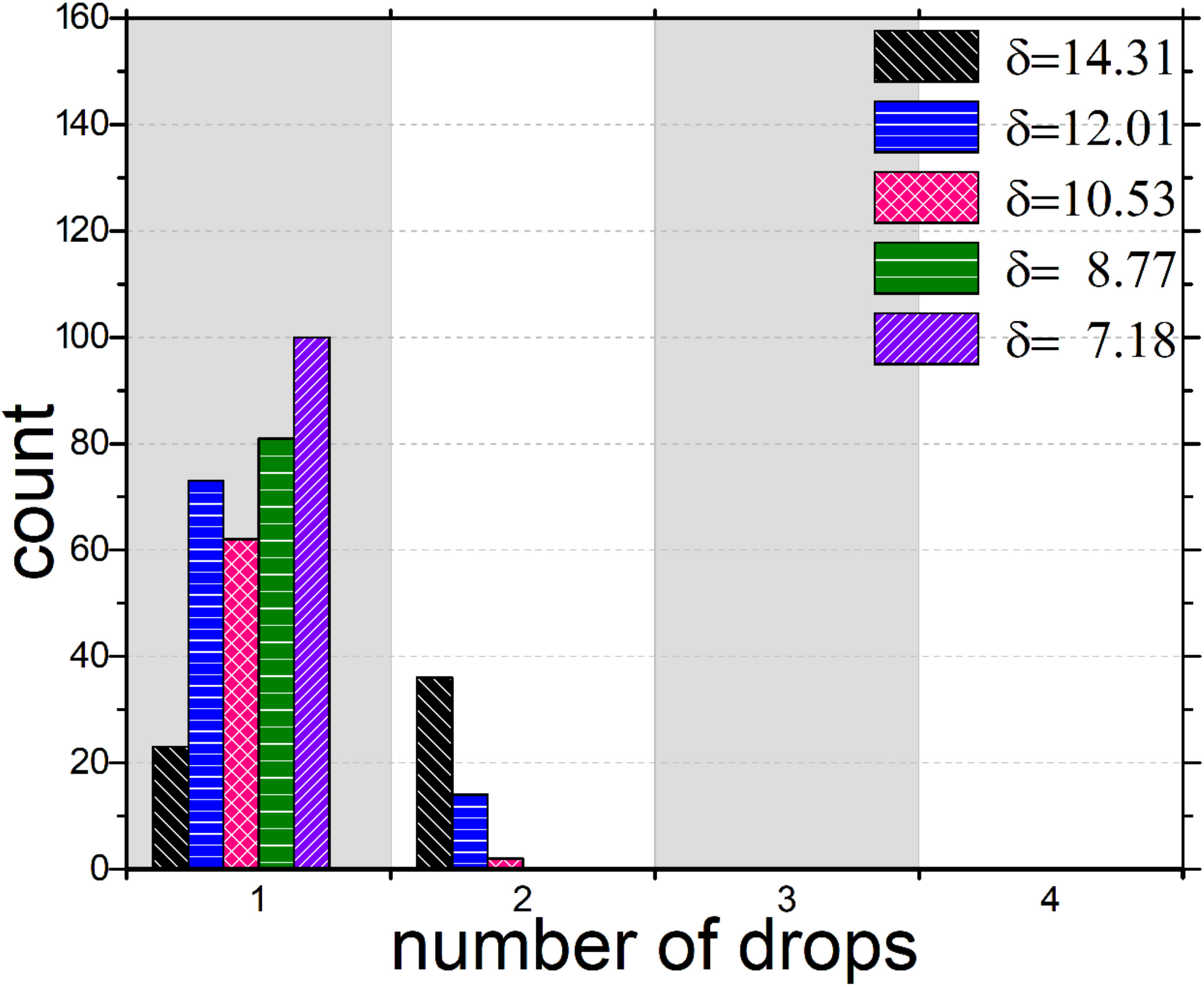}}
\caption{Number of filaments (count) that yield a certain number of drops as the thickness, $h_g$, and the aspect ratio,  $\delta=L_g/w$ are varied.}
\label{fig:drops}
\end{figure}

Figure~\ref{fig:drops} shows histograms that indicate how many times a given number of drops, $n$, (resulting from each detached filament) is observed for a given grid, characterized by the nondimensional parameter 
\begin{equation}
 \delta=\frac{L_g}{w}.
 \label{eq:delta}
\end{equation}
For instance, for the grid in Fig.~\ref{fig:grid_001}, which corresponds to $\delta=19.69$ and $h_g=10$~nm, we have three bars, namely for $2$, $3$ and $4$ drops (Fig.~\ref{fig:drops}b). However, the case with $3$ drops is far more frequent ($40$ counts) than those for $2$ and $4$ drops ($7$ and $4$ counts, respectively). So, the modal number for this grid is $n=3$. Note that each pair $(\delta,h_g)$ represents an experiment such as that in Fig.~\ref{fig:grid_001}, so that $14$ experiments are summarized in Fig.~\ref{fig:drops}. For each one, we analyze all filaments that have finished evolving; the total number can be found by adding the counts of the corresponding bars. For instance, the above mentioned case includes $51$ filaments/edges.

\begin{figure}[htb]
\includegraphics[width=0.80\linewidth]{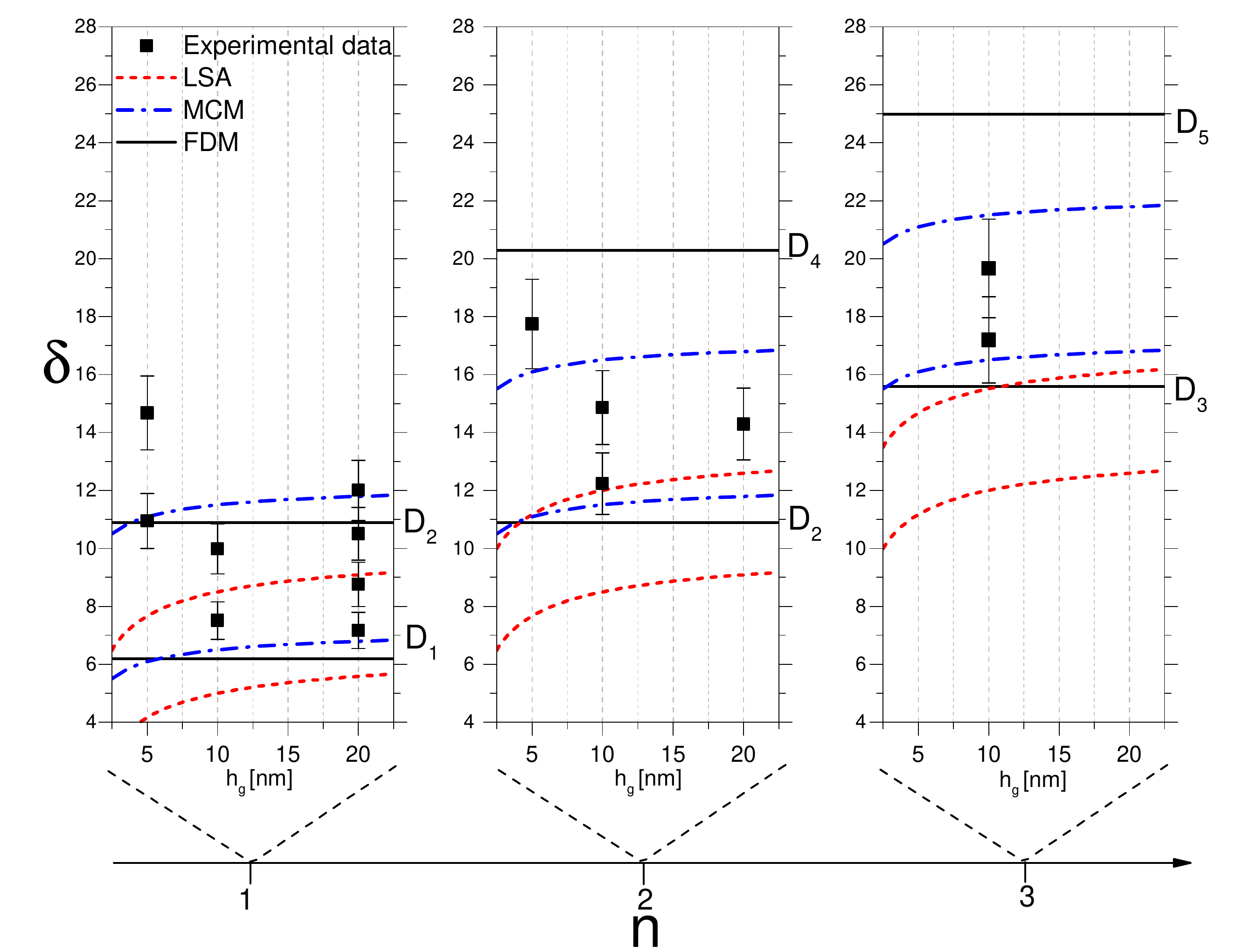}
\caption{
Each symbol corresponds to an experiment carried out with a given value of thickness, $h_g$, and the aspect ratio, $\delta=L_g/w$. The experiments are grouped according to the modal number of drops, $n$, as determined from the  histograms in Fig.~\ref{fig:drops}. The error bars correspond to the uncertainty in $w$, see Eq.~(\ref{eq:w69}). The horizontal dashed lines are the predictions obtained from linear stability analysis (LSA), showing (dimensionless) wavelength of maximum growth, $\lambda_m/w(h_g)$, see Eq.~(\ref{eq:LSA_range}). The dot--dashed curves are obtained by the Mass Conservation Model (MCM), see Eq.~(\ref{eq:delta_range}). The solid horizontal lines correspond to the Fluid Dynamic Model (FDM), see Eq.~(\ref{eq:Dk}).}
\label{fig:Delta_n}
\end{figure}

Figure~\ref{fig:Delta_n} shows the modal number obtained from the experimental results presented in Fig.~\ref{fig:drops} as $\delta$ and $h_g$ are varied. In this figure, each symbol corresponds to an experiment: for example, the experiment discussed in the preceding paragraph is represented by the point $(h_g,\delta)=(10,19.69)$ for $n=3$. The error bars correspond to uncertainty in $w$ (see Eq.~(\ref{eq:w69})). Note that each experimental point in Fig.~\ref{fig:Delta_n} corresponds to the modal value of drops from a large number of filaments (between $50$ and $300$, depending on $L_g$). The various lines shown in Fig.~\ref{fig:Delta_n} show the predictions of various models that will be discussed later in the text.   

\subsection{Linear Stability Analysis (LSA)}

As a first attempt to estimate the emerging spatial scales due to breakups of the filaments, we consider the linear stability analysis (LSA) of an infinitely long filament. There are various approaches in the literature to such an analysis; see~\cite{sekimoto_annphys87,dgk_pof09} for elaborate discussions. In particular, Fig.~5 from~\cite{dgk_pof09} compares several existing models showing that the differences between them are mostly modest for contact angles smaller than $\pi/2$. For simplicity, we consider only the results obtained by carrying out LSA within long-wave (lubrication) theory, despite the fact that the contact angles in the present problem are not small. Such LSA yields the following expression for the critical (marginal) wavenumber, $q_c$ (see Eq.~(27) in~\cite{dgk_pof09}),
\begin{equation}
 q_c \tanh(q_c/2) \tanh(1/2)=1, 
 \label{eq:lsa}
\end{equation}
where $q_c=2\pi  w/\lambda_c$, and $\lambda_c$ is the critical wavelength. The solution of this equation is $q_c=2.536$, quite differently from a straightforward Raleigh-Plateau (RP) criterion $q_c^{RP}=1$. Even if it is usual to consider RP criterion as a first rough approximation, it lacks other essential features of the problem, such as the contact line physics, and so Eq.~(\ref{eq:lsa}) is more appropriate. Then, the distance between drops for the varicose unstable mode is given by the most unstable wavelength, $\lambda_m=\sqrt{2} \lambda_c=3.504\,w$. Since all the grid sides are of the length $L_g$, one expects that $n$ is related to how many times $\lambda_m$ fits into $L_g$. Note that the experiments show that a drop is present at each corner, and then these positions must correspond to maximums of the perturbation, which in turn restricts the admissible Fourier modes of the LSA. Since $n$ accounts for the number of internal drops (in between both corner drops separated by $L_g + w_g$), we have the condition
\begin{equation}
(n+1) \lambda_m \leq L_g + w_g < (n+2)\lambda_m. 
 \label{eq:LSA_range}
\end{equation}
Note that the upper bound for $n$ drops corresponds to the lower bound for $n+1$ drops.

Figure~\ref{fig:Delta_n} shows the predicted limits for $\delta$ as dashed lines for $n=1$, $2$ and $3$. Clearly, the comparison with the experiment shows that the straightforward LSA yields a too narrow range of values of $\delta$ for given $n$, leaving some experimental points out of it.   The agreement is lacking particularly for smaller values of $\delta$, as expected since the LSA theory as presented assumes infinite filament length. Moreover, LSA is inconsistent with the the experimental evolution of the system: the breakups do not occur simultaneously, but in a cascade process starting from the filament ends. Therefore, a model that takes into account these features of the problem is required.  We proceed with discussing two of such models. 

\subsection{Mass conservation model (MCM)}

Here we consider the fluid mechanical description recently reported by Cuellar et al.~\cite{cuellar_pof17} in the context of microfluidic experiments carried out with 
the silicon oil grids.   Although the scale of the experiments considered in~\cite{cuellar_pof17} is different, visual similarity of the instabilities to the ones considered in the present paper suggests that the instability mechanism may be similar. The macroscopic experiments from~\cite{cuellar_pof17} provide however significantly more detailed information about grid evolution, that is useful in explaining the present experimental results for which such detailed information is not available.  

\begin{figure}[hbt]
\includegraphics[width=0.5\linewidth]{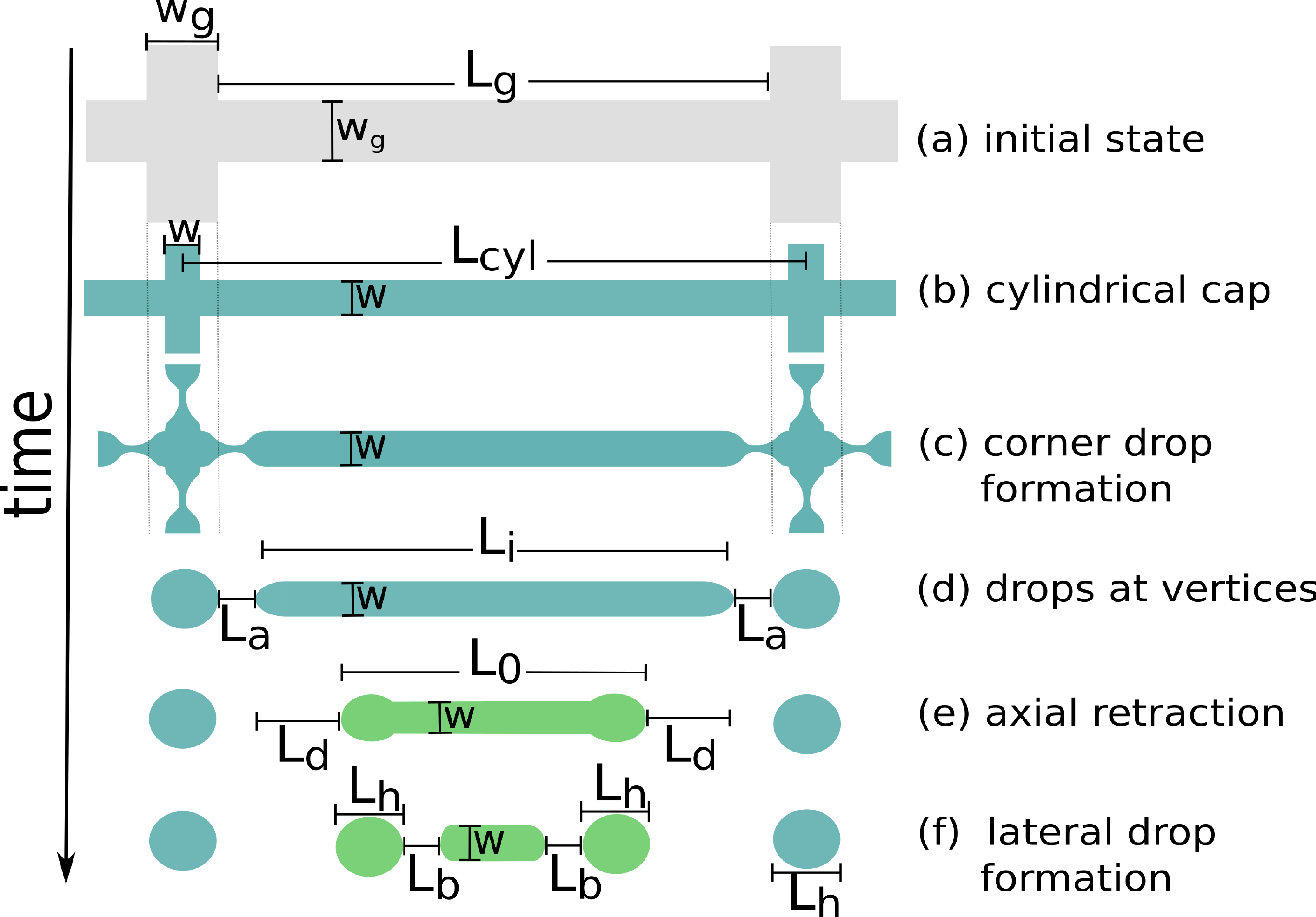}
\caption{Scheme illustrating the time evolution of a portion of square grid.  See the text for the discussion of the  various stages of breakup process.}
\label{fig:scheme}
\end{figure}

Figure~\ref{fig:scheme} illustrates the instability mechanism discussed extensively in~\cite{cuellar_pof17}; here we provide a brief overview of the main features.  The first step (between Fig.~\ref{fig:scheme}a and b), in which the rectangular cross section of the filament changes to a cylindrical one, has been discussed previously (see e.g. Eq.~(\ref{eq:width})). Afterwards, drops start developing at each grid intersection, so that cross--like structures are formed and bridge regions 
appear at their arms (see Fig.~\ref{fig:scheme}c).  This process leads to bridge ruptures and the formation of detached filaments of length $L_i$ (see Fig.~\ref{fig:scheme}d). 
The length of these bridges, $L_a$, is approximately equal to those of the arms of a cross that dewets to form a corner drop. Then, we can write
\begin{equation}
L_i=L_g-2 L_a.
\label{eq:Li}
\end{equation}

The detached filaments  retract axially  and bulged regions start forming at the ends (the stages between Fig.~\ref{fig:scheme}d and Fig.~\ref{fig:scheme}e). These bulges stop after having retracted a distance $L_d$, so that the new filament length is
\begin{equation}
L_0=L_i-2 L_d .
\label{eq:L0}
\end{equation}
By comparing Fig.~\ref{fig:scheme}e with Fig.~\ref{fig:grid_001}b, we expect that this is the stage achieved at the end of the first pulse. This expectation 
is supported by the fact that the positions of the bulges in Fig.~\ref{fig:grid_001}b are coincident with the two side drops close to those at the corners (see Fig.~\ref{fig:grid_001}c).
This is the case for most of the detached filaments observed in Fig.~\ref{fig:grid_001}b. The bulges are connected to the filament by means of additional bridges, whose
length is denoted by $L_b$.   The rupture of these bridges leads to the formation of additional drops (see Fig.~\ref{fig:scheme}f), and the breakup process continues 
until only final drops remain, as also observed in~\cite{gonzalez_07}. 

Clearly, the bridge breakup process is the key feature. When the bulges have achieved the equilibrium as in Fig.~\ref{fig:scheme}e, we assume a balance 
between the capillary pressure (due to the longitudinal and transverse curvatures) in the bulge, and that in the filament. Since both the bulge and the detached drop adopt 
approximately the shape of a spherical cap (no visible hysteresis effects are present here, in contrast to the microscopic experiments~\cite{cuellar_pof17}, the balance yields the value of the 
bulge size as (see Eqs.~(10) and (11) in~\cite{cuellar_pof17}) 
\begin{equation}
L_h=2\,w,
\label{eq:Lh}
\end{equation}
so that the drop volume is 
\begin{equation}
 V_{drop}=\frac{\pi}{6} w^3 \left(2 + \cos{\theta} \right) \sec^2{\frac{\theta}{2}} \tan{\frac{\theta}{2}}.
 \label{eq:Vdrop}
\end{equation}

Note that the filament of length $L_i$ can be thought as consisting of portions of length $d$, each one leading to the formation of a single drop. Thus, $d$ can be calculated as the ratio between the volume drop, $V_{drop}$, and the cross section of the filament, $A$:
\begin{equation}
 d= \frac{V_{drop}}{A}=\frac{8 \pi}{3} w \frac{2 + \cos{\theta}} {\theta - \cos{\theta} \sin{\theta}}
 \sin^2{\frac{\theta}{2}} \tan{\frac{\theta}{2}}.
 \label{eq:d_theta}
\end{equation}
In order to obtain $n$ drops from a filament of length $L_i$, we have 
\begin{equation}
n d \leq L_i < (n+1)d.
\label{eq:Li_range}
\end{equation}
However, the value of $L_i$ is not readily available from the experiments, and needs to be estimated. We note first that the corner drops result from dewetting of a part of the grid at the intersections (see Fig.~\ref{fig:scheme}c). This part corresponds to a cross whose arms have length $L_a$, so that its volume is $ V_{cross}= A \left(4 L_a + w_g \right) $. Assuming that the corner drops have similar size as those resulting from the filament breakup, we can write $V_{cross}=V_{drop}$, and obtain
\begin{equation}
 L_a=\frac{d-w_g}{4}.
\label{eq:La}
\end{equation}
By using this result in Eq.~(\ref{eq:Li}),  Eq.~(\ref{eq:Li_range}) gives
\begin{equation}
 \left( n + \frac{1}{2} \right) \frac{d}{w} - \frac{w_g}{2 w} \leq \delta \equiv \frac{L_i}{w}< \left(n + \frac{3}{2}\right) \frac{d}{w} - \frac{w_g}{2 w}.
\end{equation}
In particular, for $\theta=69^\circ$ Eq.~(\ref{eq:d_theta}) yields 
\begin{equation}
d=(5 \pm 0.1)w,
\label{eq:d69}
\end{equation}
and using Eq.~(\ref{eq:w69}) to include the dependence of $w$ on $h_g$, we have
\begin{equation}
 5 \left(n + \frac{1}{2} \right) - \frac{1}{4}\sqrt{\frac{w_g}{h_g}} \leq \delta < 5 \left(n + \frac{3}{2}\right) - \frac{1}{4}\sqrt{\frac{w_g}{h_g}}.
 \label{eq:delta_range}
\end{equation}

The expressions in Eq.~(\ref{eq:delta_range}) are plotted in Fig.~\ref{fig:Delta_n} as MCM--curves $\delta$ versus $h_g$ for given $n$. In general, the bounds given by MCM have a better agreement with the experiments than LSA.

\begin{figure}[htb]
\subfigure[]
{\includegraphics[width=0.45\linewidth]{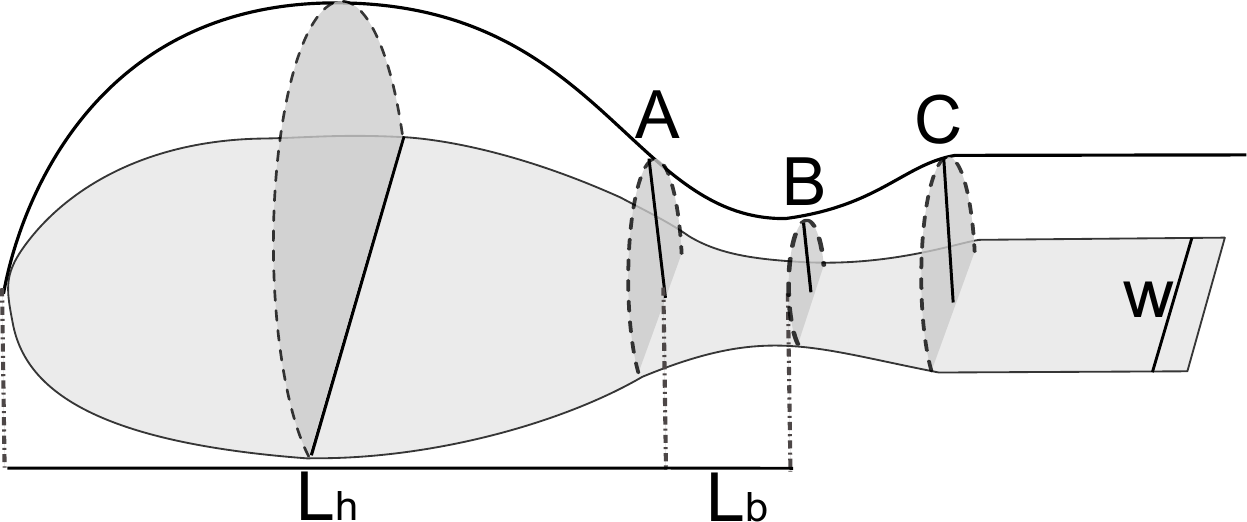}}
\subfigure[]
{\includegraphics[width=0.45\linewidth]{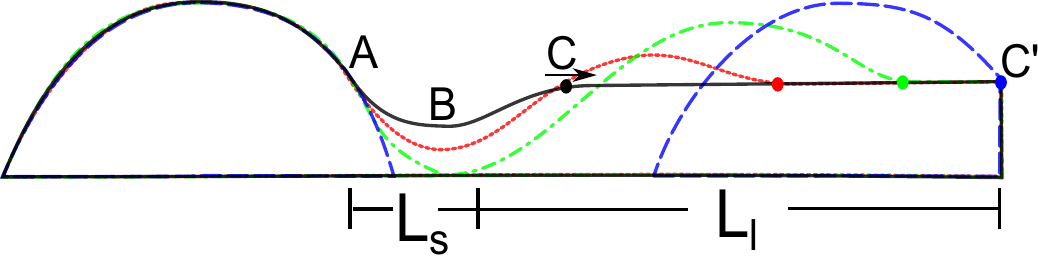}}
\caption{(a) Sketch of the head and bridge regions showing the parameters used in the model. (b) Sketch of the longitudinal section showing the interpretation of both roots for $L_b$, namely, $L_s$ and $L_l$.}
\label{fig:sketch}
\end{figure}

\subsection{Fluid dynamical model (FDM)}

Next, we consider a model that includes analysis of the flow during a breakup~\cite{cuellar_pof17}. Figure~\ref{fig:sketch} illustrates different stages that can be observed during filament evolution. First, the bulge at the filament end stops its axial retraction when it reaches a certain size, at which the bulge is at equilibrium with the unperturbed filament, see Fig.~\ref{fig:sketch}a. Let us call by A the static point where the bulge is connected to the filament. The connecting region (bridge) may develop a small disturbance in the form of a neck. As a bridge narrows, an axial Stokes flow develops there (in the next subsection we confirm that inertial effects are not of relevance here). This flow is due to the dynamic balance between the viscous forces and pressure difference between the bulge and the depressed center of the bridge (A and B in Fig.~\ref{fig:sketch}a, respectively, where $L_b$ is the distance between them). This pressure difference occurs because of the distinctive curvatures (longitudinal and transversal) at the bridge center B,  and those at its ends (A and C, only transversal), where the curvatures are equal to that of an unperturbed filament. Note that the pressure at C is the same to that at any point in the rest of the filament (e.g., C' in Fig.~\ref{fig:sketch}b). Since both the head and the unperturbed remaining filament are at equilibrium, points A, C and C' have the same pressure (note that A and C do not need to be symmetric with respect to B).  As the pressure at point B is different from that at the points A and C, there is an outflow from B that further depletes the neck region leading to an eventual breakup. Requiring a balance between the resulting Stokes flow and the pressure differences between B and the points A or C, one finds that there are two possible equilibrium distances from the center of the neck (B) to the points with the unperturbed pressure of a straight filament (say A and C'). 
 
As discussed in~\cite{cuellar_pof17}, there are two positive values of $L_b$: one for a short bridge, $L_s$, and another for a long one, $L_l$, namely 
\begin{equation}
 L_s=0.597 w, \qquad L_l=4.1 w,
 \label{eq:LsLl}
\end{equation}
(both calculated for $\theta=69^\circ$). The smaller root, $L_s$, corresponds to the distance between A and B. In order to understand the larger one, note that (as the flow develops in the neck, the point C moves  away from B towards the filament.  Simultaneously, a new bulge starts to form (dashed red line in Fig.~\ref{fig:sketch}b). 
When the breakup occurs at B, the bulge dewets and grows (dot--dashed green line in Fig.~\ref{fig:sketch}b). Finally, C stops at C', which is an equivalent point to A, because the new bulge (dotted blue line in Fig.~\ref{fig:sketch}b) is identical to the former one since it has reached the curvatures needed to be at equilibrium with the filament. Consequently, the distance between the fixed point B (breakup point) and C' (where the static bulge and filament meet) corresponds to the second root, $L_l$.

\begin{figure}[htb]
\includegraphics[width=0.6\linewidth]{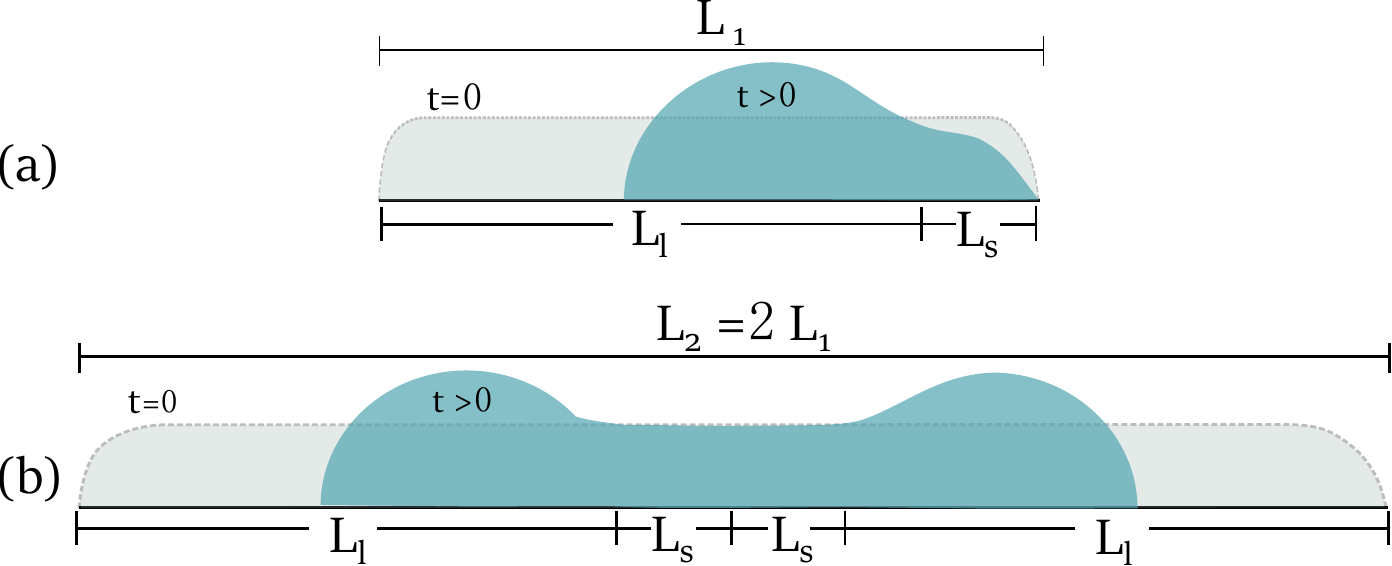}
\caption{Sketches of the filament showing the parameters used in the model to define the limiting lengths of the filaments that yield: (a) one drop, and (b) two drops.}
\label{fig:evol}
\end{figure}

Based on this interpretation of the second root, $L_l$, we can write (see Fig.~\ref{fig:scheme}e)
\begin{equation}
 L_l \approx L_d + L_h,
 \label{eq:Ll}
\end{equation}
as confirmed by the experiments in~\cite{cuellar_pof17}. Therefore, the characteristic length of a filament needed for the formation of a single drop is $L_1=L_l+L_s=4.697\, w$.  Figure \ref{fig:evol}a illustrates the introduced quantities.   Note that $L_1$ is  conceptually equivalent to the length $d$ in Eq.~(\ref{eq:d_theta}) (see also Eq.~(\ref{eq:d69})). Although $L_s$ is derived from the FDM model for the breakup of a single filament, it is close to the value found for the bridges (cross arms) that occur at the intersection of perpendicular filaments, namely $L_a$ that was obtained in the MCM. It is noteworthy that while one model focuses on the mass conservation and the other one on the dynamic effects, they yield similar results. 

Within the dynamical model, if $L_i= L_2 \equiv 2 L_1$, there is the possibility of generating two drops when the small bridge between the two heads formed from both ends of the filament is long enough to allow for a breakup at a distance $L_s$ from each static bulge (see Fig.~\ref{fig:evol}b). Following a similar reasoning, a general formula for the limits of $L_i$ that allow for the formation of $n$ drops can be written as:
\begin{equation}
L_n \equiv n L_1, \quad n=1,2,\ldots
\label{eq:Ln}
\end{equation}

Note however that these limits are only lower limits for the existence of a certain number of drops, not the upper ones. For example, when $L_i$ is slightly below $L_2$, there is the possibility that both heads coalesce into a single drop. Then, the upper limit of one drop can be estimated as $L_2$. Regarding the upper limit for more drops, this coalescence process could occur on both sides of the remaining bridge, and therefore its maximum length should be $2 L_1$. Then, the upper limit for the formation  of $n$ drops can be written as $(n+2) L_1$ for $n\geq 2$.  In order to compare this model with the experimental data in Fig.~\ref{fig:Delta_n}, we use Eqs.~(\ref{eq:delta}) 
 and (\ref{eq:Li}) to define
\begin{equation}
 D_k= \frac{k L_1 + 2 L_a}{w}, \qquad k =1,2,\ldots
 \label{eq:Dk}
\end{equation}
where 
\begin{equation}
 L_a=0.75\,w,
 \label{eq:La69}
\end{equation}
for $\theta=69^\circ$ (see Eqs.~(\ref{eq:w69}), (\ref{eq:La}) and (\ref{eq:d69})). Although the model is based on rather rough approximations, the predicted limits agree very well with the experimental data (see FDM lines in Fig.~\ref{fig:Delta_n}). These limits are horizontal lines because Eq.~(\ref{eq:Dk}) does not depend on $w(h_g)$, in contrast to LSA and MCM. Note that FDM predicts overlapping $\delta$--intervals for the existence of a certain number of drops. For instance, for $D_3<\delta<D_4$ it is possible to find either $2$ or $3$ drops, as observed in the experiments.

Summarizing, we have used three models to predict the number of drops and compared the predictions to experiments. The most accurate seems to be FDM, which takes into account the flow in the bridge regions. MCM, which is simpler, provides less accurate results, while the LSA model is the least accurate. A substantial difference between FDM and both LSA and MCM is that the former takes into account the actual sequence of events that lead to the final droplet configuration, such as the axial dewetting, the bridge formation and breakup. This iterated sequence propagates from both filament ends towards the center. On the other hand, both LSA and MCM assume simultaneous evolution of unstable varicose modes and breakups. 

The models discussed so far have not considered the time scales involved in breakup process. In the following section, we present numerical simulations that allow to discuss this time scale.   

\subsection{Numerical simulations}

In this section we discuss time dependent numerical simulations of the dewetting and breakup processes~\cite{ghigliotti_pof13,schulkes_jfm96}.  We will see that the results of these simulations are consistent with both the models and experimental results discussed so far.  Furthermore, we will see that the time scales emerging from the simulations are consistent with the experimental ones, for reasonable values of the slip length that is used to define fluid/solid boundary condition.  Although precise comparison of the time scales between experiments and simulations is difficult due to the fact that only limited amount of information is available from the experiments, we find this consistency encouraging.  

For efficiency of the computations, we consider only a single square of the grid. Since the initial dewetting process that evolves a strip from rectangular to circular cross-section is very fast, we do not simulate this process. Instead, we consider that at $t=0$, the unit cell of the grid is formed by four filaments of a cylindrical cap shape, and of the length $L_{cyl}=L_g+ w_g$ (see Fig.~\ref{fig:scheme}b).

The time evolution is obtained by numerically solving the dimensionless Navier-Stokes equation
\begin{equation}
Re \left[ \frac{\partial \vec {v}}{\partial t}+(\vec{v} \cdot \vec {\nabla} )  \vec{v} \right] 
= - \vec {\nabla} p + \nabla ^{2} \vec {v},
\label{eq:NS}
\end{equation}
where $Re=\rho \gamma w/\mu^2$ is the Reynolds number. Here, the scales for the position $\vec x=(x,y,z)$, time $t$, velocity $\vec {v}=(u,v,w)$, and pressure $p$, are the $w$, $t_c=\mu w/\gamma$, $U=\gamma/\mu$, and $\gamma/w$, respectively. The fluid parameters for the melted Ni are: density $\rho=7.905$~g/cm$^3$, viscosity $\mu=0.0461$~poise, and surface tension $\gamma=1780$~dyn/cm. For the considered experiments we have $w=80$~nm, so that $t_c=0.2$~ns and $Re=53$.  Note that this value of $Re$ is a consequence of the scaling used for the characteristic velocity, $U$, which yields a capillary number $Ca=\mu U/\gamma=1$. This is done since the value of $U$ is not known \emph {a priori}. As it will soon be seen (see e.g. the slopes in Fig.~\ref{fig:xhead}), the maximum dimensionless flow velocities are of the order of $10^{-2}$, so that the actual $Re$ and $Ca$ are much smaller than the above numbers.

The normal stress at the free surface accounts for the Laplace pressure in the form
\begin{equation}
 \Sigma_n = -\left( \vec \nabla_{\tau} \cdot \hat n \right) \hat n,
\end{equation}
where $\hat n$ and $\hat \tau$ are the surface normal and tangential vectors.  Since the surrounding fluid (e.g., air) is passive, we assume that the tangential stress is zero at this surface, i.e. $\Sigma_{\tau}=0$.

Regarding the boundary condition at the contact line, we set there a fixed contact angle, $\theta$. As it is commonly done for the problems involving moving contact lines, we relax the no slip boundary condition at the substrate through the Navier formulation (see e.g.~\cite{Haley91}),
\begin{equation}
 v_{x,y} = \ell \, \frac{\partial v_{x,y}}{\partial z}\quad \hbox{at $z=0$,} 
 \label{eq:slip}
\end{equation}
where $\ell$ is the slip length. Based on the experimental comparison from the previous work~\cite{wu_lang11} considering evolution of liquid metals, we use the value of slip length 
$\ell = 20$~nm; this choice is discussed further below.

We use a Finite Element technique in a domain which deforms with the moving fluid interface by using the Arbitrary Lagrangian-Eulerian (ALE) formulation~\cite{hughes_cmame81,donea_cmame82,christodoulou_cmame92,hirt_cmame97}. The interface displacement is smoothly propagated throughout the domain mesh using the Winslow smoothing algorithm, which consists of mapping an isotropic grid in computational space onto an arbitrary domain in physical space, and it is usually more effective than the Laplace smoothing approach~\cite{winslow_jcp66,knupp_ec99,harakh_jcp97}.  The main advantage of this technique is that the fluid interface is and remains sharp~\cite{tezduyar_cmame06}, while its main drawback is that the mesh connectivity must remain the same, which precludes achieving situations with a topology change (e.g., when the filaments break up). The default mesh used is unstructured, and consists typically of $13\times 10^4$ triangular elements and $4\times 10^4$ tetrahedral elements.

\begin{figure}[htb]
     \centering
     \subfigure[$\, t=40, \quad t_c=8$~ns]
	{\includegraphics[width=0.3\linewidth]{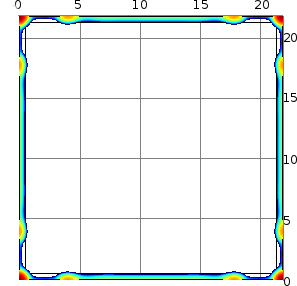}}
     \subfigure[$\, t=80, \quad t_c=16$~ns]
        {\includegraphics[width=0.3\linewidth]{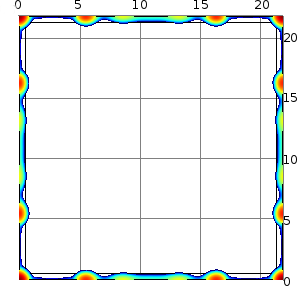}}
     \subfigure[$\, t=105, \quad t_c=21$~ns]
	{\includegraphics[width=0.3\linewidth]{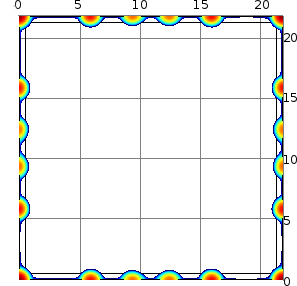}}
     \caption{Time evolution of the fluid thickness for the parameters of Fig.~\ref{fig:grid_001} using $\ell=20$~nm. Here, we have $w=80$~nm and $L_{cyl}=1749$~nm. The lengths are in units of $w$.}
     \label{fig:square}
\end{figure}

Since the problem is symmetric with respect to the axis of each filament, we consider only the interior of the square, and apply symmetry boundary conditions along its four sides.  Figure~\ref{fig:square} shows the time evolution of this square for  the parameters as in Fig.~\ref{fig:grid_001} where $\delta=19.69$. The three stages correspond to: (a) the breakups that lead to the corner drops and the formation of the detached filaments with the bulges at their ends, (b) the first breakups of these bulges, and (c) the final configuration with four drops along the grid side.

\begin{figure}[htb]
     \centering
      \includegraphics[width=0.7\linewidth]{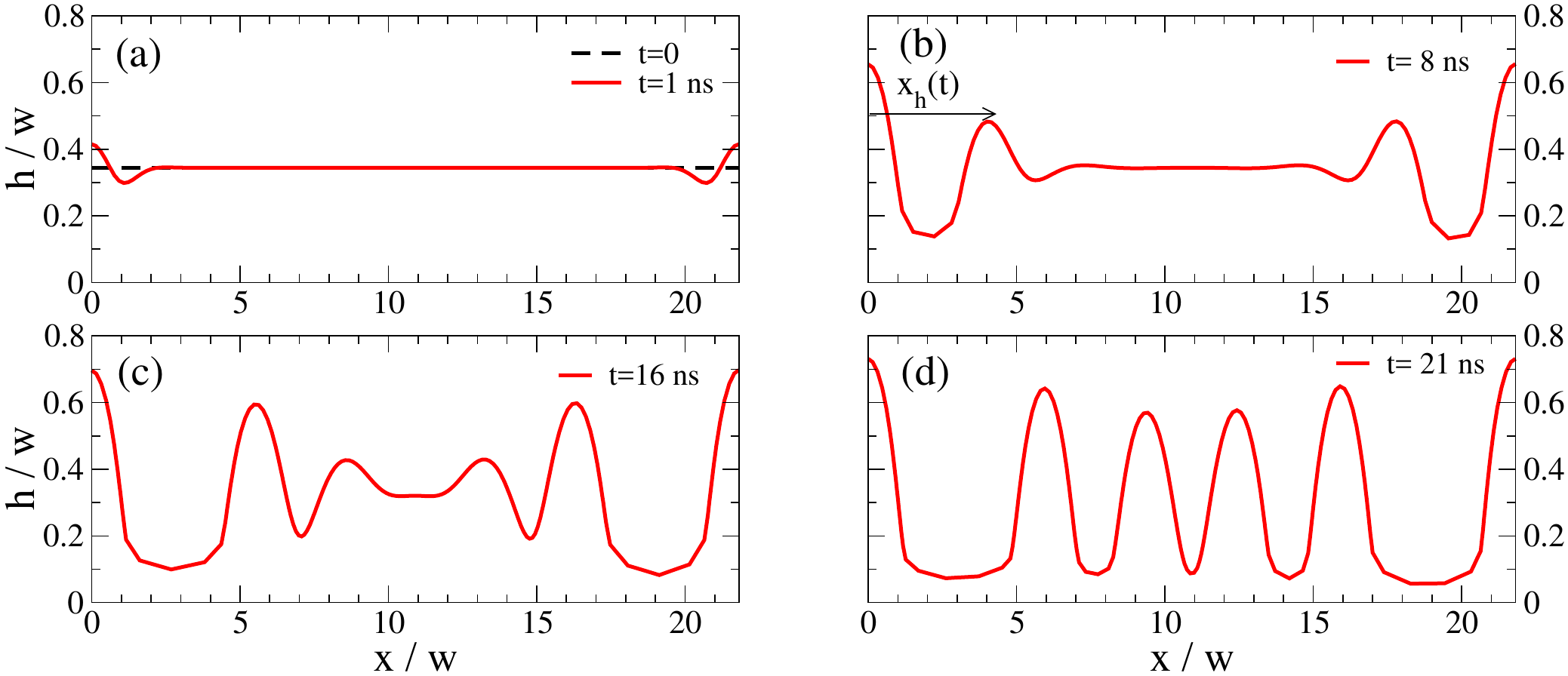}
     \caption{Thickness, $h$, along a filament ($x$-coordinate is defined along the symmetry line of a filament), at different times for the grid shown in 
     Fig.~\ref{fig:square} with $\ell=20$~nm: (a) $t=0$ and $1$~ns, (b) $t=8$~ns, (c) $t=16$~ns, (d) $t=21$~ns. 
     Note the corner drops at $x=0$ and $x/w=L_{cyl}=21.81$. The arrow indicates the position of maximum thickness at the bulk, $x_h(t)$, 
     further discussed in the text.  
      }
     \label{fig:hx_t}
\end{figure}

Figure~\ref{fig:hx_t} shows the thickness profile along the symmetry line of one of the sides of the square. Since complete breakup cannot be simulated using the present numerical method, a remnant  film remains between the drops. Note that this particular case leads to four drops, consistently with some of the experimental outcomes (see Figs.~\ref{fig:drops}b and \ref{fig:grid_001}c), although most of the time this geometry leads to three drops (modal number $n=3$). We have also observed in the simulations that cases with slightly smaller $\delta$ lead to three drops. Moreover, a comparison of the simulations with FDM shows that this particular case is in the overlapping interval between $n=3$ and $n=4$.

\begin{figure}[hbt]
\subfigure[\,$L_g=1387$~nm ($\delta=17.21$)]
{\includegraphics[width=0.325\linewidth]{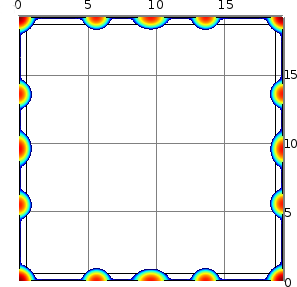}}
\subfigure[\,$L_g=987$~nm ($\delta=12.25$)]
{\includegraphics[width=0.325\linewidth]{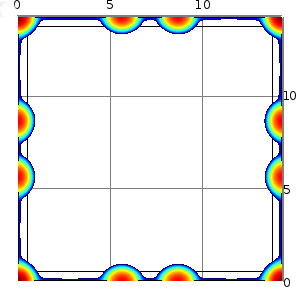}}
\subfigure[\,$L_g=606$~nm ($\delta=7.52$)]
{\includegraphics[width=0.325\linewidth]{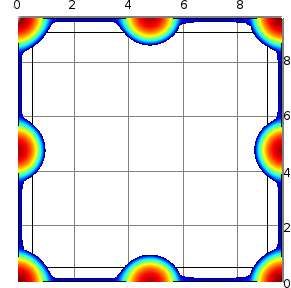}}
\caption{Final numerical drop patterns for the experimental cases shown in Fig.~\ref{fig:grid_Lg} using $\ell=20$~nm.  The results are shown at the late times 
such that no further evolution is expected.}
\label{fig:numgrid_Lg}
\end{figure}

\begin{figure}[htb]
     \centering
     \includegraphics[width=0.4\linewidth]{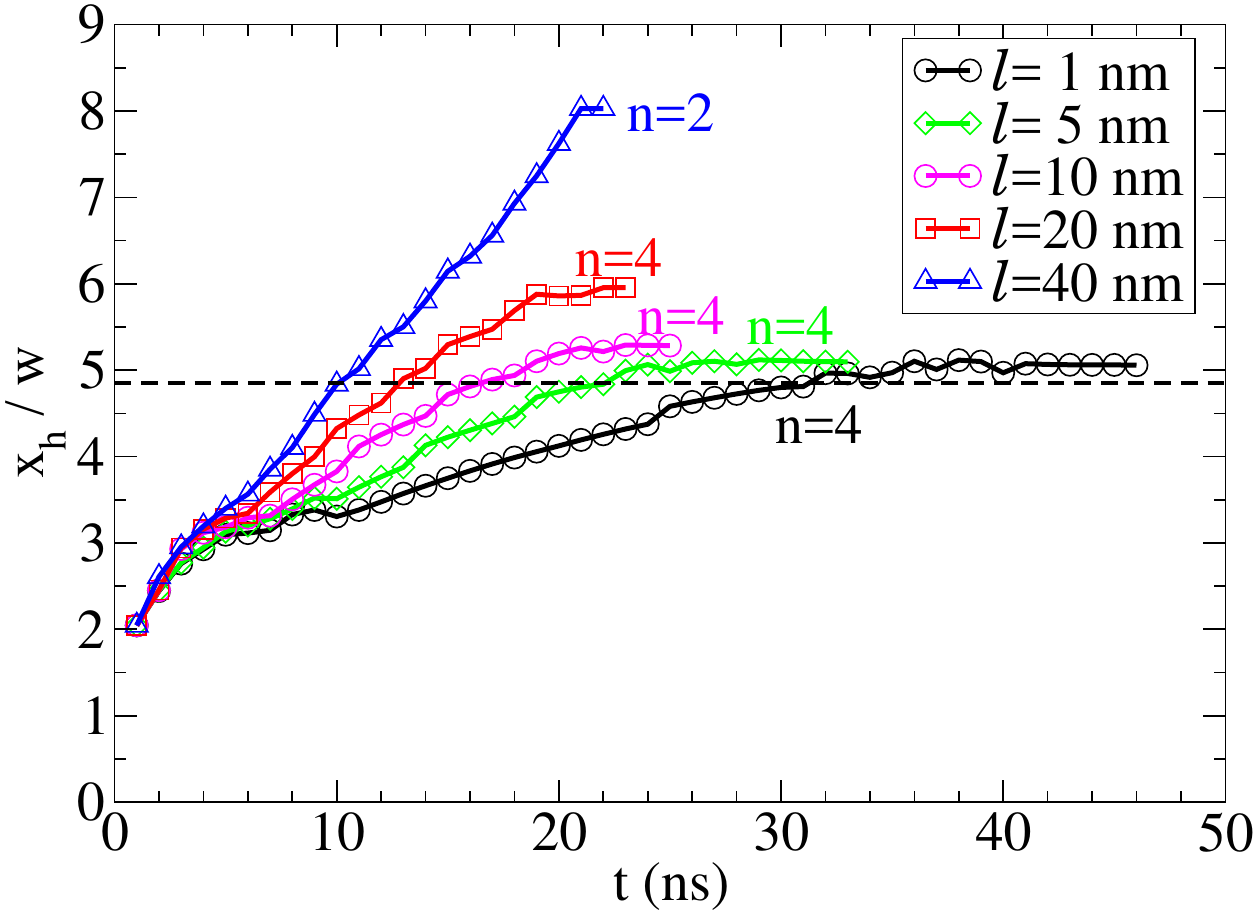}
     \caption{Time evolution of the position of the point of maximum height in the bulge, $x_{h}$, as a function of time for several values of $\ell$. The $x$--coordinate is measured from the filament
     intersection and $n$ stands for the number of drops formed along each side of the square. The horizontal dashed line stands for $x_h^{max}$ as predicted by FDM.}
     \label{fig:xhead}
\end{figure}

Figure~\ref{fig:numgrid_Lg} shows the results for the parameters corresponding to the experiments from Fig.~\ref{fig:grid_Lg}, illustrating how a decrease of $\delta$ also leads to increased number of drops in the simulations. Moreover, in these cases the obtained value of $n$ is fully consistent with the experiments.


Regarding the choice of the slip length, $\ell = 20$~nm, we note that the main expected influence of the value of $\ell$ is on the time scale of the  evolution. To discuss this issue further, we carry out simulations with $\ell$  in the range $[1,40]$~nm, and record the position of the maximum height in the bulge as a function of time. Figure~\ref{fig:xhead} shows the corresponding results, together with the resulting number of drops. Not only the time scale of the problem is affected by $\ell$ (smaller $\ell$ implies slower evolution), but also the final value of $x_{h}$ changes and, eventually, the resulting number of drops as well.   While precise comparison of the time scales between experiments and simulations is not possible since we do not know exactly when the 
evolution stops in the experiments, it is encouraging to find comparable time scales between experiments and simulations for a reasonable value of slip length. 

We are also in position to compare the simulation results with the models considered so far. Figure~\ref{fig:xhead} shows (dashed line) the position of the maximum thickness at the bulge (see Fig.~\ref{fig:scheme}e) 
\begin{equation}
 x_{h}^{max}= L_a + L_d + L_h,
\end{equation}
where $L_a$ is given by Eq.~(\ref{eq:La69}). We can estimate $x_{h}^{max}$ by resorting to the FDM. According to  Eqs.~(\ref{eq:LsLl}) and (\ref{eq:Ll}) we find $x_{h}^{max} \approx 4.85\,w$, shown as the dashed line in Fig.~\ref{fig:xhead}.  As a consequence, the values of $\ell$ that lead to a dewetting distance of the filament end that are in agreement with the model are in the interval $(1,20)$~nm. Moreover, assuming that the bulge is at rest at the end of first pulse ($t \approx 18$~ns), we consider that $\ell=20$~nm is an appropriate choice to account for the experimental data, consistently with the previous works~\cite{wu_lang11} that considered similar type of experiments. 

\subsection{Further effects at the intersections}

\begin{figure}[htb]
     \centering
     \includegraphics[width=0.35\linewidth]{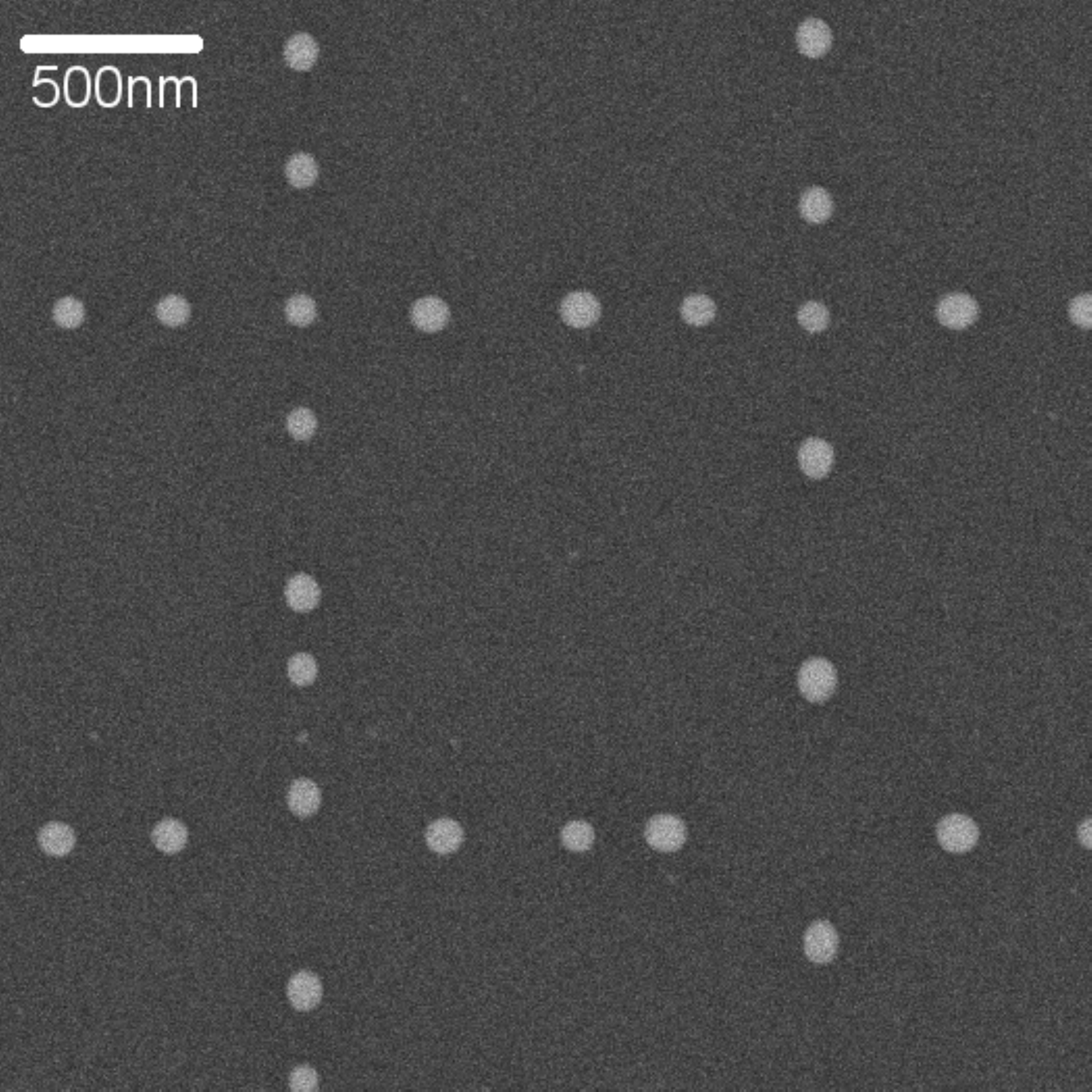}
     \caption{Closeup of a SEM for a grid with $h_g=5$~nm and $L_g=1556$~nm ($\delta=24.4$). Note that the lower right corner drop is missing.}
     \label{fig:vert_drops}
\end{figure}

We observe in the experimental pictures that there are some cases where no drop is formed at the vertices (see Fig.~\ref{fig:vert_drops}). This anomalous effect is more frequent for smaller values of $h_g$, e.g. $h_g=5$~nm in  Fig.~\ref{fig:grid_Lg}c. One explanation for such behavior is an increased importance of the initial irregularities of the strip thickness, leading to instabilities of the free surface rather than those related to the contact line. Consequently, the position of the bridges could be altered by other mechanisms, which could be more of a local character and less related to the symmetry of the system. Such anomalous  behavior is particularly common for $\delta=14.65$ for $h_g=5$~nm and $n=1$, and it is therefore not surprising that this particular data point in Fig.~\ref{fig:Delta_n} seems to be an outlier which does not agree with the proposed models. Careful analysis of the data shows that for $\approx 37 \%$ of the vertices, a corner drop is missing for this particular geometry.    

\begin{figure}[htb]
     \centering
     \subfigure []
     {\includegraphics[width=0.35\linewidth]{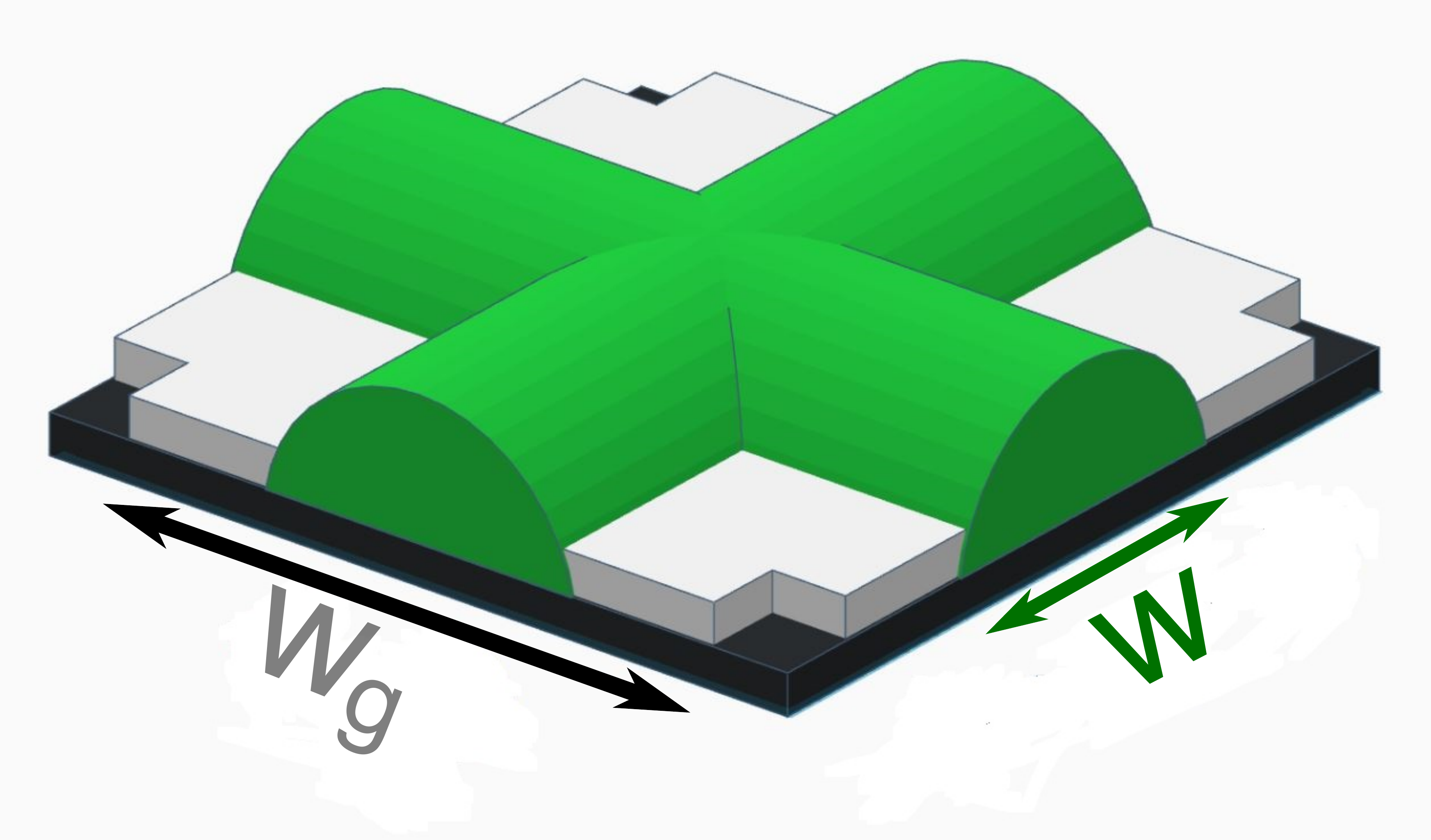}}
     \subfigure []
     {\includegraphics[width=0.4\linewidth]{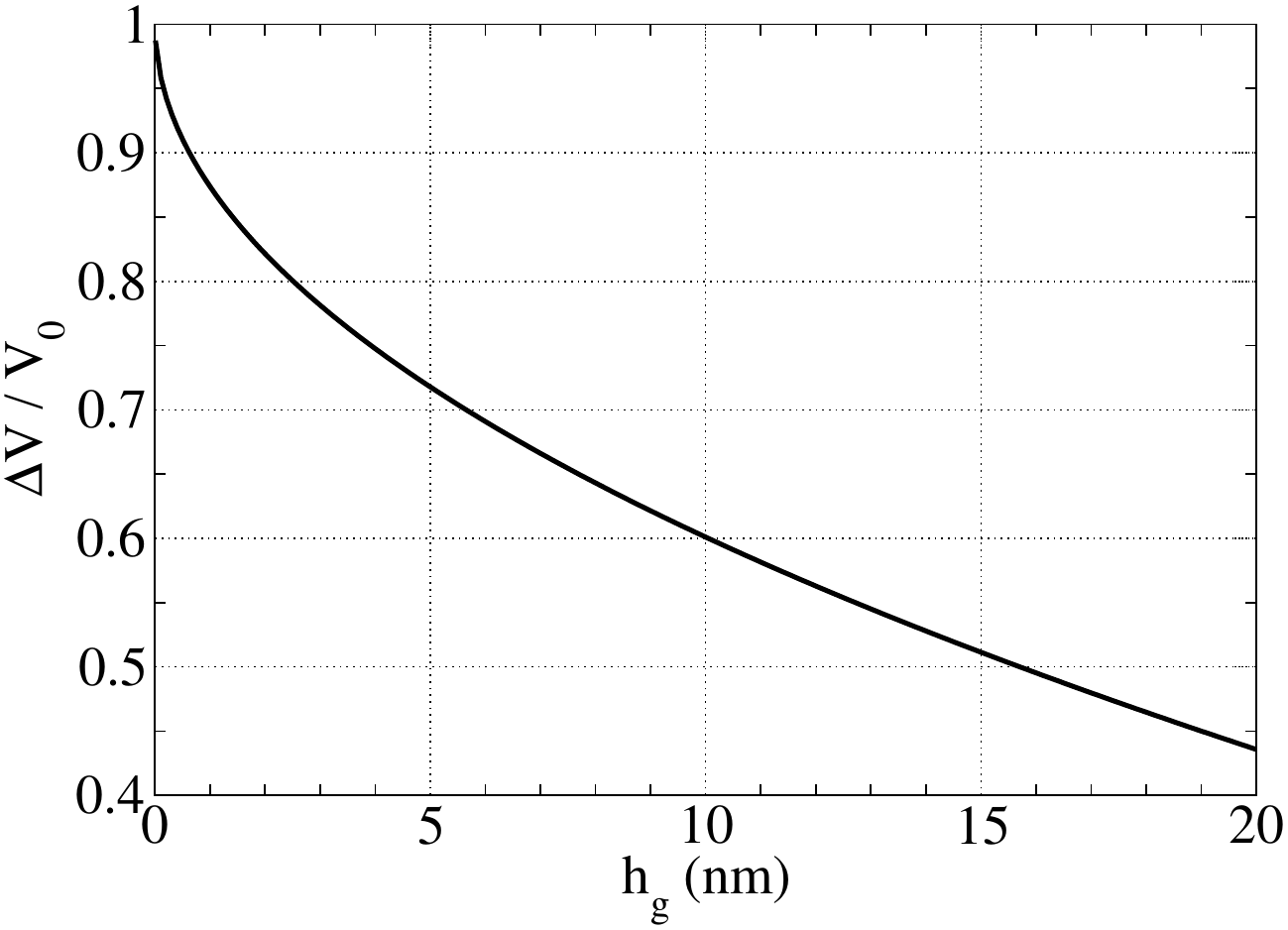}}
     \caption{(a) Sketch showing the intersection region for the original grid (dashed lines) of width $w_g$ with rectangular cross section of thickness $h_g$, and the (liquefied) cylindrical cap filaments of width $w$. $V_0$ corresponds to the square with thick lines, and $V_{cross}$ to the colored cross region. The dashed diagonals stand for the intersections of the cylindrical surfaces. (b) Relative variation of volume in the intersection region as a function of $h_g$ (see Eq.~(\ref{eq:DeltaV})) when comparing the original volume $V_0$ with the assumed cylindrical cap arms as depicted in (a).}
     \label{fig:Dvol}
\end{figure}

We can rationalize this effect by noting that there is a volume difference in the vertex region between the original intersection of two strips with rectangular transversal section and the assumed cross with cylindrical cap arms after the fast initial dewetting stage (see Figs.~\ref{fig:scheme}a and b). In fact, the volume of the original intersection region, $V_0=h_g w_g^2$, must be compared with the volume, $V_{cross}$, of the cross region with cylindrical transversal section and width $w$ (see Fig.~\ref{fig:Dvol}a). Thus, the relative variation can be calculated as
\begin{equation}
 \frac {\Delta V}{V_0} = \frac{V_{cross} - V_0}{V_0}= \frac {w^2}{6 h_g w_g^2} \left[ 3 (w - w_g) \cot \theta + 2 w \cot^3 \theta + (3 w_g \theta \sin \theta - 2 w ) \csc^3 \theta \right] - 1,
 \label{eq:DeltaV}
\end{equation}
which is plotted in Fig.~\ref{fig:Dvol}b as a function of $h_g$ for $\theta=69^\circ$, $w_g=160$~nm and $w$ as given by Eq.~(\ref{eq:w69}). Note that this difference can be as large as $\approx 0.7$ for $h_g=5$~nm, while it reduces significantly for larger values of $h_g$, such as $h_g=10$ or $20$~nm. 

This volume deficit in the experiments may be the reason why the corner drop is frequently  missing for small values of $h_g$.  Such a deficit implies the formation of either necks at the cross arms or a depression at the cross center. In general, the first option is more likely to happen, but the probability of the second one increases as $h_g$ decreases, since $\Delta V$ is so large that neck formation is not enough to compensate for it. For larger $h_g$, this effect appears to be less relevant, since no missing corner drops are observed for $h_g=10$ and $20$~nm,  and only necks in the arms are formed. 

\section{Summary and Conclusions}

In this work we report and analyze a series of experiments focusing on the formation of two--dimensional drop patterns by carrying out pulsed laser--induced dewetting (PliD) of a square grid of Ni strips on silicon wafers. By means of well established nanofabrication techniques we are able to precisely control the initial far from equilibrium geometry and the liquid lifetime via nanosecond laser melting. The results are presented as a series of snapshots (SEM's) which show the grid evolution as the number of pulses is increased, and they are interpreted in terms of fluid mechanical models of increasing complexity intended to predict the number of drops that will result from the breakup. The models predictions are given by the curves in Fig.~\ref{fig:Delta_n}, which are compared with the data from $14$ experiments for different thicknesses $h_g$.

The advantage of this type of experiments is that they allow to study not only the two--dimensional structure of the grid as a whole, but also two other fundamental problems, namely, the formation of the corner drops as well as the dewetting and breakup of short filaments. The modeling of these two phenomena has been combined with the analysis of the experimental grid patterns. Moreover, the whole grid structure provides a large number of intersections and filaments ($50$ or more) under identical conditions, which is very convenient to verify repeatability and perform statistical analysis.

The most basic approach is to use the results of the LSA for an infinitely long filament under long--wave approximation. However, this attempt seems to be too crude for the present problem since its predictions do not compare well with the data.   We expect that the lack of agreement comes from the assumption of infinitely long filament, and not from the use of long--wave approximation which is known to produce accurate results 
in this particular context of filament breakup even for large contact angles~\cite{dgk_pof09}.   
A better approximation is obtained by resorting to a detailed mass conservation formulation that assumes that all drops can be represented by spherical caps. Finally, we obtain an even better agreement with the experiments by applying a fluid dynamical model (FDM), which was previously successful to account for similar experiments on microscopic scale~\cite{cuellar_pof17}, and that takes into account the dynamics of the filament breakup.

The time evolution of the grid is also numerically simulated by solving the full Navier--Stokes equation assuming a  fixed contact angle and a given slip length, $\ell$. In general, the numerical results regarding the final number of drops along each side of the grid agree with both experiments and model. By analyzing the position of the maximum at the bulge, we obtain that $\ell$ should be around $20$~nm to obtain times scales comparable to those in the experiments.  This value of slip length is consistent with earlier work~\cite{wu_lang11}, that also considered
the evolution of liquid metals of nanoscale thickness.

While the agreement between relatively simple models, simulations of Navier-Stokes equations, and experiments, is promising, we note that additional effects could be relevant in the context of dewetting of liquid metal filaments (and other geometries) on nanoscale, such as thermal effects in the metal and substrate~\cite{ajaev_pof03}, as well as the phase change processes. Studies that will include some of these effects are left for the future work.  

\begin{acknowledgments}
I. Cuellar and P. Ravazzoli acknowledge post--graduate student fellowships from Consejo Nacional de Investigaciones Cient\'{\i}ficas y T\'ecnicas (CONICET, Argentina). J. Diez and A. Gonz\'alez acknowledge support from Agencia Nacional de Promoci\'on Cient\'{\i}fica y Tecnol\'ogica (ANPCyT, Argentina) with grant PICT 1067/2016. P. Rack acknowledges support from NSF CBET grant 1603780. The experiments and the lithographic patterning were conducted at the Center for Nanophase Materials Sciences, which is a DOE Office of Science User Facility. L. Kondic acknowledges support by NSF CBET grant 1604351.
\end{acknowledgments}

\end{document}